\documentclass[english]{aastex63}
\usepackage[T1]{fontenc}
\usepackage[latin9]{inputenc}
\usepackage{bm}
\usepackage{amstext}
\usepackage{graphicx}

\makeatletter
\usepackage{savesym}
\savesymbol{iint}
\savesymbol{iiint}
\savesymbol{iiiint}
\usepackage{amsmath}
\usepackage{array}
\usepackage{multirow}
\usepackage[caption=false]{subfig}

\makeatother

\usepackage{babel}
\begin{document}
\global\long\def\grad{\bm{\nabla}}%
 
\global\long\def\sgshj#1{Y_{jm}^{#1}}%
\global\long\def\sgshl#1{Y_{\ell m}^{#1}}%
 
\global\long\def\vgshj#1{\mathbf{Y}_{jm}^{#1}}%
 
\global\long\def\shl{Y_{\ell m}}%
 
\global\long\def\shj{Y_{jm}}%
 
\global\long\def\gfnomega#1#2#3{G_{#2,#3\omega}^{#1}}%
 
\global\long\def\gfnlomega#1#2{G_{#2,\ell\omega}^{#1}}%
 
\global\long\def\vgshl#1{\mathbf{Y}_{\ell m}^{#1}}%

\title{A general formulation for computing spherical helioseismic sensitivity
kernels while incorporating systematical effects}

\author[0000-0001-6433-6038]{Jishnu Bhattacharya}
\affiliation{Center for Space Science, New York University Abu Dhabi, Abu Dhabi, P.O. Box 129188, UAE}
\author[0000-0003-2896-1471]{Shravan M. Hanasoge}
\affiliation{Department of Astronomy and Astrophysics, Tata Institute of Fundamental Research, Mumbai - 400005, India}
\affiliation{Center for Space Science, New York University Abu Dhabi, Abu Dhabi, P.O. Box 129188, UAE}
\author{Katepalli R. Sreenivasan}
\affiliation{Center for Space Science, New York University Abu Dhabi, Abu Dhabi, P.O. Box 129188, UAE}
\affiliation{New York University, NY, USA 10012}

\begin{abstract}
As helioseismology matures and turns into a precision science, modeling finite-frequency, geometric and systematical effects is becoming increasingly important. Here we introduce a general formulation for treating perturbations of arbitrary tensor rank in spherical geometry using fundamental ideas of quantum mechanics and their extensions in geophysics. We include line-of-sight projections and center-to-limb differences in line-formation heights in our analysis. We demonstrate the technique by computing a travel-time sensitivity kernel for sound-speed perturbations. The analysis produces the spherical harmonic coefficients of the sensitivity kernels, which leads to better-posed and computationally efficient inverse problems.
\end{abstract}

\section{Introduction}

Helioseismology has been successful at producing an isotropic, radially
stratified and nearly adiabatic model of the solar interior that fits
the observed eigenfrequencies of standing modes in the Sun to a high
accuracy. The close match facilitates further use of seismic waves
as a diagnostic tool within the ambit of perturbation theory. Subsurface
inhomogeneities inside the Sun leave their imprint on the propagation
of seismic waves, and a measurement of the differences induced may
be traced back to the inhomogeneity through an inverse problem. In time-distance helioseismology, a widely used technique, wave travel-times, measured as a function various angular distances on the surface of the
Sun, are used to infer the solar subsurface \citep{1993Natur.362..430D}.
A perturbation to the solar interior --- for example in the form
of an altered sound-speed profile or the presence of a flow --- would
change the speed at which waves propagate, and as a result would lead
to observable differences in the time taken by waves to travel between
two points on the Sun. A physically consistent theory is necessary
to relate these travel-time shifts to subsurface inhomogeneities,
one that in turn may be used to set up an inverse problem to infer
their profiles and magnitudes. Such a problem is usually posed in
the form of an integral relation, where the wave travel-time between
two points on the solar surface is related to subsurface features
through a response function that is referred to as the sensitivity
kernel. The sensitivity kernel encapsulates the physics of the wave
propagation as well as the measurement procedure, and serves as a
map from subsurface model parameter to the surface observables. As
an extension it also enables us to carry out the inverse problem,
that of using the surface measurements to infer the subsurface structure
and dynamics of the Sun.

Inverse problems in the domain of time-distance helioseismology have
to overcome several distinct challenges --- (1) realization noise limits the inferential ability, (2) ill-posed inverse problems where
a large number of sub-surface models may match the surface observations
leading to non-uniqueness of solutions, and (3) simplifications
in the physical model of wave propagation that are invoked in order to reduce computational complexity. Addressing the issue of measurement
noise involves carrying out appropriate averages over the data to
improve the signal strength. We focus on the two other aspects in
this work, firstly that of wave physics and secondly that of the inverse
problem being well-posed. A common approach that has been used in
the past is the JWKB approximation, also referred to as ``ray theory\textquotedbl, 
where it is assumed that the deviations to the isotropic model are
much smoother compared to the wavelengths of the seismic waves and
they only impact the waves through a locally altered wavelength \citep{1996ApJ...461L..55K,1997ASSL..225..241K,2004ApJ...603..776Z}.
This assumption is expected to fall short when applied to spatially
localized features on the Sun such as sunspots or supergranular flows
whose sizes are comparable to the first Fresnel zone \citep{2001ApJ...561L.229B}.
An improvement to the ray-theory may be obtained through the first
Born approximation, where the assumption of scale-separation between
the seismic wavelength and the structural inhomogeneity is lifted.
The impact of the perturbation on the waves is computed in the single-scattering
approximation which is expected to hold for small-amplitude changes
to the solar background. \citet{2002ApJ...571..966G} expound on the mathematical underpinnings of a finite-frequency sensitivity kernel computation, and since then such calculations have been carried out both semi-analytically by summing over normal modes by \citet{2007AN....328..228B},
and numerically by \citet{2007AN....328..319H,Gizon2017A&A...600A..35G}. This approach
proceeds by assuming homogeneity or isotropy in the underlying solar
model, allowing for the use of known functional forms of the eigenfunctions in computing 
the kernels. The numerical approach is more general as it does not require symmetries in the underlying medium, although it might use such properties for computational efficiency. The present work falls in the semi-analytical category and computes kernels about an isotropic model of the Sun.

While perturbations to the wave propagation are addressed by incorporating single-scattering theory, there remains  the aspect of correctly modeling the geometry of the domain.
The Sun is spherical, and analysis of large-scale features must therefore account for spherical geometry firstly through their spatial functional form, and secondly through systematics in
the measurement procedure such as line-of-sight projection and center-to-limb
variation in spectral-line-formation heights. There are other systematic
effects such as foreshortening and convective blueshift that affect
seismic measurements, but we do not address these in the current work.
Several authors including \citet{2016ApJ...824...49B,2017ApJ...842...89M,2018A&A...616A.156F}
have developed mathematical approaches to compute Born sensitivity
kernels in spherical geometry, where \citet{2016ApJ...824...49B}
and \citet{2017ApJ...842...89M} have computed the three-dimensional
spatially varying kernels and \citet{2018A&A...616A.156F} have computed spherical-harmonic coefficients of the kernel. Both \citet{2016ApJ...824...49B}
and \citet{2017ApJ...842...89M} have ignored the impact of line-of-sight
projection by assuming the direction of projection to be radial, whereas
\citet{2018A&A...616A.156F} have avoided it by modeling the divergence
of the wave velocity. The approach by \citet{2018A&A...616A.156F}
has the advantage of sparsity, as an analysis of large-scale features
on the Sun may be restricted to low-degree spherical-harmonic modes,
which reduces the number of free parameters significantly. The work
that we present in this paper is similar in spirit to \citet{2018A&A...616A.156F},
as we illustrate a procedure to compute sensitivity kernels for spherical-harmonic components of the parameters being sought. We demonstrate that it is possible to carry out the analysis in terms of the vector wave displacement without making the aforementioned assumptions about the observations. This provides the extra advantage of incorporating line-of-sight projection effects seamlessly into
our analysis by computing the appropriately directed components of
 wave velocity. This analysis rests on the use of vector spherical
harmonics --- rank-$1$ analogs of scalar spherical harmonics ---
that serve as complete bases with which to expand vector fields on a sphere. The analysis of vector spherical harmonics is well known in the fields
of quantum mechanics \citep{1988qtam.book.....V} and geophysics \citep{1976RSPTA.281..195J,DahlenTromp},
and has been used previously in the field of helioseismology by various authors including \citet{1991ApJ...369..557R} and \citet{2000SoPh..192..193B}
to expand oscillation eigenfunctions in the Sun. We borrow much of
this mathematical framework and apply it in the context of computing sensitivity kernels. This work therefore has to be seen as an incremental effort to improve upon pre-existing techniques to compute sensitivity kernels through the use of established mathematical constructs, and not as an effort to develop the formalism of generalized spherical harmonics in the context of seismology. The technique that emerges has the added advantage that it is independent of the tensor rank of the parameter of interest, so the same analysis procedure may be used to compute sensitivity kernels for scalar quantities such as sound speed, vector fields such as flow velocity or rank-$2$ fields such as the Maxwell stress tensor. In this work we apply the technique to compute kernels for sound-speed perturbations in order to highlight the mathematical technique without being overly burdened by algebraic complexity.

Time-distance helioseismology has historically used the travel time as
a single parameter derived from surface-wave measurements that
is fit through perturbative improvements in the background model.
Recently, \citet{2017A&A...599A.111N} have extended this to include
the mismatch between wave amplitudes as a second parameter, one that
may be fit independently of wave travel times. The difference between
the sensitivity kernels in these two approaches is only in a weight
function that projects the mismatch in measured cross-covariances
to the parameter. Our approach may therefore be seamlessly extended
to compute amplitude kernels as well, which gets us one step closer
to full-waveform inversions including the entire measured wave field. As we demonstrate in this paper, the predominant impact of including line-of-sight projection is in the amplitude of the cross-covariances, therefore the analysis presented here is perhaps more relevant in the context of full-waveform inversions as opposed to travel-times. Nevertheless, as we are able to model the change in cross-covariance directly, we may also obtain any projected parameter that we might seek to fit.

The paper is arranged in various sections that develop the analysis incrementally. In section \ref{sec:GSH} we review the specific spherical harmonics that we use in our analysis. This is followed by section \ref{sec:Green-function} where we compute the Green function and its change in the first Born approximation in presence of a sound-speed perturbation. In section \ref{sec:Synthetic-cross-covariances} we compute the resulting change in the line-of-sight projected wave cross-covariance using the Green function from section \ref{sec:Green-function}, and finally in section \ref{sec:kernel} we compute the travel-time sensitivity kernel using the cross-covariance from section \ref{sec:Synthetic-cross-covariances}.

\section{Generalized spherical harmonics\label{sec:GSH}}

\subsection{Helicity basis}

The analysis of functions in spherical geometry is usually carried out
in the spherical polar coordinates $\left(r,\theta,\phi\right)$, where $r$ is radius, $\theta$ is co-latitude and $\phi$ is longitude.
The direction of increase in each coordinate is denoted by the corresponding
unit vectors $\mathbf{e}_{r}$, $\mathbf{e}_{\theta}$ and $\mathbf{e}_{\phi}$.
The analysis of spherical harmonics may be simplified by switching
over to a different basis --- one obtained as a linear combination
of the spherical polar unit vectors --- defined as 
\begin{align}
\mathbf{e}_{+1} & =-\frac{1}{\sqrt{2}}\left(\mathbf{e}_{\theta}+i\mathbf{e}_{\phi}\right),\label{eq:helicity_spherical}\\
\mathbf{e}_{0} & =\mathbf{e}_{r},\nonumber \\
\mathbf{e}_{-1} & =\frac{1}{\sqrt{2}}\left(\mathbf{e}_{\theta}-i\mathbf{e}_{\phi}\right).\nonumber 
\end{align}
We refer to this basis as the `helicity basis' following \citet{1988qtam.book.....V}.
The helicity basis vectors satisfy $\mathbf{e}_{\mu}=\left(-1\right)^{\mu}\mathbf{e}_{-\mu}^{*}$.
We note that these are the covariant basis vectors, and the corresponding
contravariant ones are defined as $\mathbf{e}^{\mu}=\mathbf{e}_{\mu}^{*}$.
These vectors satisfy the orthogonality relation $\mathbf{e}^{\nu}\cdot\mathbf{e}_{\mu}=\mathbf{e}_{\nu}^{*}\cdot\mathbf{e}_{\mu}=\delta_{\mu\nu}$.

A vector may be expanded in the helicity basis as $\mathbf{v}=v^{\alpha}\mathbf{e}_{\alpha}$,
where we invoke the Einstein summation convention. The components
$v^{\alpha}$ are the contravariant components of the vector in the
helicity basis. The same vector may also be expanded in the contravariant
basis as $\mathbf{v}=v_{\alpha}\mathbf{e}^{\alpha}$, where the covariant
components $v_{\alpha}$ are related to the contravariant components
$v^{\alpha}$ through $v_{\alpha}=v^{\alpha*}$. The inner product
of two vectors may therefore be expressed as $\mathbf{v}\cdot\mathbf{w}=v_{\alpha}w^{\alpha}=v^{\alpha*}w^{\alpha}$.
The components $v^{\alpha}$ of a vector $\mathbf{v}$ transform under
complex conjugation as $v^{\alpha*}=\left(-1\right)^{\alpha}v^{-\alpha}$.

\subsection{Vector Spherical Harmonics}

Spherical harmonics form a complete, orthonormal basis with which to decompose
scalar functions on a sphere. Analogously, vector spherical harmonics
(VSH) form a basis with which to decompose vector fields on a sphere. While spherical
harmonics are uniquely defined (barring choices of phase,) vector spherical
harmonics have a leeway in this regard as we may construct various
linear combinations that act as complete bases. The specific choice
in a particular scenario depends on the geometry and symmetries inherent
in the problem. Vector spherical harmonics have been used in helioseismology
by \citet{1991ApJ...369..557R,Christensen-Dalsgaard97lecturenotes,2000SoPh..192..193B,2017MNRAS.470.2780H}
to decompose wave eigenfunctions as 
\begin{equation}
\bm{\xi}_{n\ell m}\left(\mathbf{x}\right)=\xi_{r,n\ell m}\left(r\right)Y_{\ell m}\left(\theta,\phi\right)\mathbf{e}_{r}+\xi_{h,n\ell m}\left(r\right)\grad_{\Omega}Y_{\ell m}\left(\theta,\phi\right),\label{eq:xi_VSH}
\end{equation}
where $\grad_{\Omega}$ is the covariant angular derivative on a sphere, defined
as 
\[
\grad_{\Omega}=\mathbf{e}_{\theta}\partial_{\theta}+\mathbf{e}_{\phi}\frac{1}{\sin\theta}\partial_{\phi}.
\]
The vectors fields $Y_{\ell m}\left(\theta,\phi\right)\mathbf{e}_{r}$
and $\grad_{\Omega}Y_{\ell m}\left(\theta,\phi\right)$ in this case
act as the necessary basis, albeit unnormalized. These form two of
the three components of a complete basis, the other component being
directed along the cross product of the two. We refer to this basis
--- including normalization --- as the Hansen VSH basis, in light
of their use by \citet{PhysRev.47.139}, although it is important to acknowledge that
they have also been referred to as the Chandrasekhar-Kendall basis,
following their application in the analysis of force-free magnetic
fields by \citet{1957ApJ...126..457C}. We follow the notation used
by \citet{1988qtam.book.....V} and denote the three vector fields
as 
\begin{align}
\vgshl{\left(-1\right)}\left(\hat{n}\right) & =\shl\left(\hat{n}\right)\mathbf{e}_{r},\nonumber \\
\vgshl{\left(0\right)}\left(\hat{n}\right) & =\frac{1}{\sqrt{\ell\left(\ell+1\right)}}\mathbf{e}_{r}\times\grad_{\Omega}\shl\left(\hat{n}\right),\\
\vgshl{\left(1\right)}\left(\hat{n}\right) & =\frac{1}{\sqrt{\ell\left(\ell+1\right)}}\grad_{\Omega}\shl\left(\hat{n}\right),\nonumber 
\end{align}
where $\vgshl{\left(-1\right)}$ and $\vgshl{\left(1\right)}$ are
spheroidal and $\vgshl{\left(0\right)}$ is toroidal in nature. The
eigenfunction $\bm{\xi}_{nlm}\left(\mathbf{x}\right)$ is strictly
spheroidal, therefore it lacks a component along $\vgshl{\left(0\right)}$.

We obtain a second useful basis through linear combinations of the
Hansen VSH as
\begin{equation}
\begin{aligned}\vgshl 1 & =\frac{1}{\sqrt{2}}\left(\vgshl{\left(1\right)}-\vgshl{\left(0\right)}\right),\\
\vgshl{-1} & =\frac{1}{\sqrt{2}}\left(\vgshl{\left(1\right)}+\vgshl{\left(0\right)}\right),\\
\vgshl 0 & =\vgshl{\left(-1\right)},
\end{aligned}
\label{eq:PB_Hansen_conversion}
\end{equation}
The two bases are related through a rotation by $45^{\circ}$ about
the radial direction $\mathbf{e}_r$ at each point. We refer to this basis
as the Phinney-Burridge (PB) VSH following their use by \citet{1973GeoJ...34..451P}.
The PB VSH satisfy 
\begin{equation}
\vgshl{\alpha}\left(\hat{n}\right)=Y_{\ell m}^{\alpha}\left(\hat{n}\right)\mathbf{e}_{\alpha},\label{eq:PB_helicity}
\end{equation}
where $Y_{\ell m}^{\alpha}\left(\hat{n}\right)=\sqrt{\left(2\ell+1\right)/4\pi}\,d_{-\alpha,-m}^{\ell}\left(\theta\right)e^{im\phi},$
and $d_{m,n}^{\ell}\left(\theta\right)$ is an element of the Wigner
d-matrix. We follow the terminology of \citet{DahlenTromp} and refer
to the scalar component $Y_{\ell m}^{\alpha}\left(\hat{n}\right)$
as a generalized spherical harmonic. Owing to this diagonal nature
of the PB VSH in the helicity basis, an expansion of a vector field
in the PB VSH basis may equivalently be thought of as an expansion
in the helicity basis. This serves as the bridge between the components
of a vector field in the Hansen VSH basis and those in the spherical
polar one, the steps involved being: (a) expand the field in the Hansen
VSH basis, (b) transform to the PB VSH basis using Equation \eqref{eq:PB_Hansen_conversion},
(c) transform to the helicity basis using Equation \eqref{eq:PB_helicity},
and finally (d) transform to the spherical polar basis using Equation
\eqref{eq:helicity_spherical}.

\subsection{Bipolar spherical harmonics}

Spherical harmonics are used to separate angular variables from
radial ones in functions of one position vector. Analogously, there
are generalizations that may be used as a complete basis with which to expand
functions of multiple position vectors. In this work we focus on two-point
functions, as the wave cross-covariance --- and as an extension the
wave travel time --- depend on two position coordinates. The angular
part of a two-point function may be represented in terms of bipolar
spherical harmonics, defined as 
\begin{align}
\sgshl{\ell_{1}\ell_{2}}\left(\hat{n}_{1},\hat{n}_{2}\right)&= \sum_{m_{1}=-\ell_{1}}^{\ell_{1}}\sum_{m_{2}=-\ell_{2}}^{\ell_{2}}C_{\ell_{1}m_{1}\ell_{2}m_{2}}^{\ell m} \times \notag \\ 
& Y_{\ell_{1}m_{1}}\left(\hat{n}_{1}\right)Y_{\ell_{2}m_{2}}\left(\hat{n}_{2}\right),
\end{align}
where $C_{\ell_{1}m_{1}\ell_{2}m_{2}}^{\ell m}$ are Clebsch-Gordan
coefficients that act as matrix elements in the transformation from
the monopolar product basis to the bipolar basis. We may extend this
coupling of modes to vector spherical harmonics and obtain a bipolar
vector spherical harmonic that is a rank$-2$ quantity.

\section{Green functions\label{sec:Green-function}}

\begin{figure}
\includegraphics[scale=0.56]{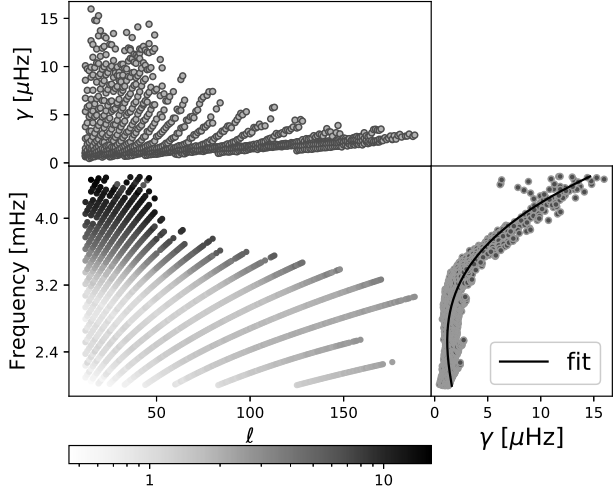}

\caption{\label{fig:damping}Model capturing seismic wave attenuation in the Sun.
The solid line in the lower right panel represents a cubic-polynomial
fit to the frequency variation of the line-width.}

\end{figure}

Seismic waves in the Sun are modeled as small-amplitude fluctuations
about a spherically-symmetric, radially stratified background structure.
Features such as inhomogeneities in the background thermal structure,
convective flows and magnetic fields are deviations from spherical
symmetry, and are treated as perturbations to the background. Accurate
modeling of seismic observables as measured at the solar surface therefore
has to account for the interaction of seismic waves with background
anisotropies, and this analysis is typically carried out in the single-scattering
first Born approximation which is valid if perturbations are small
in magnitude and their spatial scale is not significantly shorter
than the wavelength. Such an analysis proceeds by noting that the
propagation of waves is governed by the following wave equation in temporal frequency domain 
\begin{equation}
\mathcal{L}\left(\mathbf{x},\omega\right)\bm{\xi}\left(\mathbf{x},\omega\right)=\mathbf{S}\left(\mathbf{x},\omega\right),
\end{equation}
where $\mathcal{L}$ is the wave operator, $\bm{\xi}$ is the wave
displacement and $\mathbf{S}$ denotes the source that is exciting
seismic waves. The wave operator $\mathcal{L}$ is written in terms of model parameters such as density, sound speed,
convective flows and magnetic fields, each of which contributes towards
the restoring force that sustains the oscillations. As a first approximation,
we leave out advection due to convective flows and forces arising from magnetic fields and limit ourselves to
\begin{align}
\mathcal{L}\bm{\xi} =& -\rho\omega^{2}\bm{\xi}-2i\omega\gamma\bm{\xi} \notag \\ - & \grad\left(\rho c^{2}\grad\cdot\bm{\xi}-\rho\bm{\xi}\cdot\mathbf{e}_{r}g\right)-g\mathbf{e}_{r}\grad\cdot\left(\rho\bm{\xi}\right).
\end{align}
We choose a simplified model of the damping $\gamma$ by fitting a
third-order polynomial in frequency to the measured seismic
line-widths obtained from the $72$-day MDI mode-parameter set \citep{1999ApJ...523L.181S}, plotted in Figure \ref{fig:damping}. We use Model S \citep{1996Sci...272.1286C}
to obtain the structure parameters.

Waves propagating in the frequency channel $\nu=\omega/2\pi$ from source location $\mathbf{x}_{\text{src}}$ to the observation point
$\mathbf{x}$ may equivalently be described by the Green function $\mathbf{G}\left(\mathbf{x},\mathbf{x}_{\text{src}},\omega\right)$
corresponding to the wave operator $\mathcal{L}$. The Green function
in this case is a rank-$2$ tensor that satisfies 
\begin{equation}
\mathcal{L}\mathbf{G}\left(\mathbf{x},\mathbf{x}_{\text{src}},\omega\right)=\delta\left(\mathbf{x}-\mathbf{x}_{\text{src}}\right)\mathbf{I},\label{eq:lG_Delta}
\end{equation}
where $\mathbf{I}$ is the $3\times3$ identity matrix. The Green function $\mathbf{G}$ has nine components that relate the
three vector components of the source $\mathbf{S}$ to those of the
displacement $\bm{\xi}$,
\begin{equation}
\xi_{i}\left(\mathbf{x},\omega\right)=\int d\mathbf{x}_{\text{src}}G_{ij}\left(\mathbf{x},\mathbf{x}_{\text{src}},\omega\right)S_{j}\left(\mathbf{x}_{\text{src}},\omega\right).\label{eq:xi_S}
\end{equation}
The fact that eigenfunctions of the
wave operator $\mathcal{L}$ are strictly spheroidal implies that
the Green function may be expanded in terms of $\vgshj{\left(-1\right)}$
and $\vgshj{\left(+1\right)}$, and that only four of the nine components
of the Green function are independent --- the ones that relate the
spheroidal component of the sources to those of the displacement.
We use the fact that the Green function may be expressed as an outer
product of the eigenfunctions. This suggests a natural expansion of
the the Green function in the Hansen VSH basis as
\begin{align}
\mathbf{G}\left(\mathbf{x}_{\text{obs}},\mathbf{x}_{\text{src}};\omega\right)=& \sum_{\alpha,\beta=\pm1}\sum_{jm}\gfnomega{\left(\alpha\right)}{\left(\beta\right)}j\left(r_{\text{obs}},r_{\text{src}}\right)\times \notag \\ & \vgshj{\left(\alpha\right)}\left(\hat{n}_{\text{obs}}\right)\mathbf{Y}_{jm}^{\left(\beta\right)*}\left(\hat{n}_{\text{src}}\right),\label{eq:G_HansenVSH}
\end{align}
where the component $G_{\left(\beta\right),j\omega}^{\left(\alpha\right)}\left(r_{\text{obs}},r_{\text{src}}\right)$
relates the source component $S_{jm}^{\left(\beta\right)}\left(r_{\text{src}}\right)$
to the displacement component $\xi_{jm}^{\left(\alpha\right)}\left(r_{\text{obs}}\right)$
that is measured at the point $\mathbf{x}_{\text{obs}}$. The placement
of Greek indices in the super and subscripts, i.e. $\gfnomega{\left(\alpha\right)}{\left(\beta\right)}j$,
indicates that the Green function is a mixed tensor, something that might be verified by noting that $\mathbf{Y}_{jm}^{\left(\beta\right)*}=\mathbf{Y}_{jm}^{\left(\beta\right)\dagger}$ is a contravariant basis vector whereas $\vgshj{\left(\alpha\right)}$ is a covariant one. The radius $r_{\text{obs}}$
may be thought of as a representative height that arises from the
Dopplergram response function convolved with the line-formation heights.
An accurate estimate of the observation height might be necessary
for a consistent interpretation of seismic travel-times, including
its center-to-limb variation \citep{Wachter2008,Fleck2011SoPh..271...27F,2015ApJ...808...59K}.

We compute the radial components of the Green function using a high-order
finite-difference approximation to Equation \eqref{eq:lG_Delta} following
\citet{2017ApJ...842...89M}. We detail the analysis in Appendix \ref{sec:Green-functions-radial}.
The computation of line-of-sight projected measurements is easier
in the basis of the PB VSH owing to the fact that these are diagonal
in the helicity basis. We express the Green function as 
\begin{align}
\mathbf{G}\left(\mathbf{x}_{\text{obs}},\mathbf{x}_{\text{src}};\omega\right) =& \sum_{\alpha,\beta=-1}^{1}\sum_{jm}\gfnomega{\alpha}{\beta}j\left(r_{\text{obs}},r_{\text{src}}\right)\times \notag \\ & \vgshj{\alpha}\left(\hat{n}_{\text{obs}}\right)\mathbf{Y}_{jm}^{\beta*}\left(\hat{n}_{\text{src}}\right),\label{eq:G_PB}
\end{align}
where the independent components of the Green function are related
to those in the Hansen basis through
\begin{equation}
\begin{aligned}\gfnomega 00j & =\gfnomega{\left(-1\right)}{\left(-1\right)}j,\quad\gfnomega +0j=\frac{1}{\sqrt{2}}\gfnomega{\left(1\right)}{\left(-1\right)}j,\\
\gfnomega 0+j & =\frac{1}{\sqrt{2}}\gfnomega{\left(-1\right)}{\left(1\right)}j,\quad\gfnomega ++j=\frac{1}{2}\gfnomega{\left(1\right)}{\left(1\right)}j,
\end{aligned}
\end{equation}
The choice of independent $\alpha$ and $\beta$ in the PB VSH basis
is arbitrary, and we may choose to retain any four independent components
from Equation \eqref{eq:spheroidal_symmetry}.

\subsection{Change in Green function}

We choose to look at small-amplitude time-invariant perturbations
in the sound speed $c\left(\mathbf{x}\right)$, where $\mathbf{x}$
ranges over all spatial locations within the Sun where the sound speed
is finite. A local change in the sound speed --- denoted by by $\delta c\left(\mathbf{x}\right)$
--- leads to a change in the wave operator by the amount $\delta\mathcal{L}=-2\grad\left(\rho c\delta c\grad\cdot\right)$
to the linear order. A change in the wave operator by $\delta\mathcal{L}$
leads to a corresponding change in the Green function that may be
computed in the First Born approximation as
\begin{equation}
\delta\mathbf{G}\left(\mathbf{x}_{\text{obs}},\mathbf{x}_{\text{src}},\omega\right) = -\int d\mathbf{x}\,\mathbf{G}\left(\mathbf{x}_{\text{obs}},\mathbf{x},\omega\right)\cdot\delta\mathcal{L}\mathbf{G}\left(\mathbf{x},\mathbf{x}_{\text{src}},\omega\right). \label{eq:G_firstborn}
\end{equation}
We substitute $\delta\mathcal{L}=-2\grad\left(\rho c\delta c\grad\cdot\right)$
into Equation \eqref{eq:G_firstborn} and integrate by parts using
Gauss' divergence theorem. We assume zero-Dirichlet boundary conditions
in radius and therefore drop the boundary term, and use the seismic reciprocity
relation $\mathbf{G}^{T}\left(\mathbf{x}_{1},\mathbf{x}_{2}\right)=\mathbf{G}\left(\mathbf{x}_{2},\mathbf{x}_{1}\right)$, where the tensor transpose is defined as ${\bf G}^T_{ij} = {\bf G}_{ji}$,
to obtain
\begin{equation}
\delta\mathbf{G}\left(\mathbf{x}_{\text{obs}},\mathbf{x}_{\text{src}},\omega\right)=-2\int d\mathbf{x}\,\rho c\,\grad\cdot\mathbf{G}\left(\mathbf{x},\mathbf{x}_{\text{obs}},\omega\right) \grad\cdot\mathbf{G}\left(\mathbf{x},\mathbf{x}_{\text{src}},\omega\right)\delta c\left(\mathbf{x}\right).\label{eq:G_deltac}
\end{equation}
We have not explored the ramifications of including transparent boundary
conditions as used by \citet{Gizon2017A&A...600A..35G}, which lets
high-frequency waves (above the acoustic cutoff of $5.5\,\text{mHz}$)
leak out, and is expected to be a more realistic representation of
the conditions that exist on the Sun \citep[see e. g.][]{1990LNP...367...87K}. 

The perturbation $\delta c\left(\mathbf{x}\right)$ is a scalar field,
and it may be expanded in the basis of spherical harmonics as 
\begin{equation}
\delta c\left(\mathbf{x}\right)=\sum_{\ell m}\delta c_{\ell m}\left(r\right)Y_{\ell m}\left(\hat{n}\right).\label{eq:deltac_ylm}
\end{equation}
Since $\delta c\left(\mathbf{x}\right)$ is real, the spherical harmonic components $\delta c_{\ell m}\left(r\right)$
satisfy $\delta c_{\ell,-m}\left(r\right)=\left(-1\right)^{m}\delta c_{\ell m}^{*}\left(r\right)$ and therefore the components
for $m\geq0$ suffice to completely describe the spatial variation.
We seek to express Equation~\eqref{eq:G_deltac} in terms of the components
$\delta c_{\ell m}\left(r\right)$. We note that the divergence of
the Green function may be evaluated in the PB VSH basis \citep[see][]{DahlenTromp}
to
\begin{equation}
\grad\cdot\mathbf{G}\left(\mathbf{x},\mathbf{x}_{i};\omega\right)=\sum_{jm\beta}\left[\grad\cdot\mathbf{G}\right]_{\beta j\omega}\left(r,r_{i}\right)Y_{jm}\left(\hat{n}\right)\mathbf{Y}_{jm}^{\beta*}\left(\hat{n}_{i}\right),\label{eq:divG_HansenVSH}
\end{equation}
where $\left[\grad\cdot\mathbf{G}\right]_{\beta j\omega}\left(r,r_{i}\right)$
is the radial profile of the divergence of $\mathbf{G}$ for a source
directed along $\vgshj{\beta}\left(\hat{n}_{i}\right)$, and is related
to the components of the Green function through
\begin{align}
\left[\grad\cdot\mathbf{G}\right]_{\beta j\omega}\left(r,r_{i}\right) =& -\frac{\sqrt{2j\left(j+1\right)}}{r}\gfnomega 1{\beta}j\left(r,r_{i}\right)+ \notag \\ & \left(\frac{d}{dr}+\frac{2}{r}\right)\gfnomega 0{\beta}j\left(r,r_{i}\right).\label{eq:divG_radial}
\end{align}
Substituting Equations \eqref{eq:deltac_ylm} and \eqref{eq:divG_HansenVSH}
into Equation \eqref{eq:G_deltac} and integrating over the scattering
angle $\hat{n}$, we obtain
\begin{align}
\delta\mathbf{G}\left(\mathbf{x}_{\text{obs}},\mathbf{x}_{\text{src}},\omega\right) & =-2\int r^{2}dr\,\rho c\sum_{\ell m}\delta c_{\ell m}\left(r\right) \notag \times \\ & \sum_{\beta_{1},\beta_{2}=\pm1}\sum_{j_{1}m_{1}}\sum_{j_{2}m_{2}}\left[\grad\cdot\mathbf{G}\right]_{\beta_{1}j_{1}\omega}\left(r,r_{\text{obs}}\right)\times \notag \\ & \left[\grad\cdot\mathbf{G}\right]_{\beta_{2}j_{2}\omega}\left(r,r_{\text{src}}\right)\times \notag\\
 & \sqrt{\frac{\left(2j_{1}+1\right)\left(2j_{2}+1\right)\left(2\ell+1\right)}{4\pi}} \times \notag \\ & \left(\begin{array}{ccc}
j_{1} & j_{2} & \ell\\
0 & 0 & 0
\end{array}\right)\left(\begin{array}{ccc}
j_{1} & j_{2} & \ell\\
m_{1} & m_{2} & m
\end{array}\right)\times \notag \\ & \mathbf{Y}_{j_{1}m_{1}}^{\beta_{1}*}\left(\hat{n}_{\text{obs}}\right)\mathbf{Y}_{j_{2}m_{2}}^{\beta_{2}*}\left(\hat{n}_{\text{src}}\right),\label{eq:deltaG_int_angle}
\end{align}
where the angular degrees $j_{1}$, $j_{2}$ and $\ell$ are related
by the triangle constraint $\left|j_{1}-j_{2}\right|\leq\ell\leq j_{1}+j_{2}$,
and the Wigner $3$-j symbols are non-zero only for $m_{1}+m_{2}+m=0$.
For a particular choice of $\ell$, we choose $j_{1}$ to vary
independently, which would constrain $j_{2}$ to the range $\left|j_{1}-\ell\right|\leq j_{2}\leq j_{1}+\ell$.
We may also choose $m_{1}$ to vary independently for a particular
choice of $m$, which would peg $m_{2}$ to $m-m_{1}$. This reduces
the number of terms contributing to Equation \eqref{eq:deltaG_int_angle}
significantly. In subsequent analysis, we use Clebsch-Gordan coefficients
instead of the Wigner $3$-j symbols, the two being related through
\[
C_{j_{1}m_{1}j_{2}m_{2}}^{\ell m}=\left(-1\right)^{j_{1}-j_{2}+m}\sqrt{2\ell+1}\left(\begin{array}{ccc}
j_{1} & j_{2} & \ell\\
m_{1} & m_{2} & -m
\end{array}\right).
\]
The motivation for this switch is that we use bipolar spherical harmonics,
and Clebsch-Gordan coefficients are the matrix elements corresponding
to the transformation between the bipolar and the product of monopolar
harmonics.

\begin{figure*}
\includegraphics[scale=0.75]{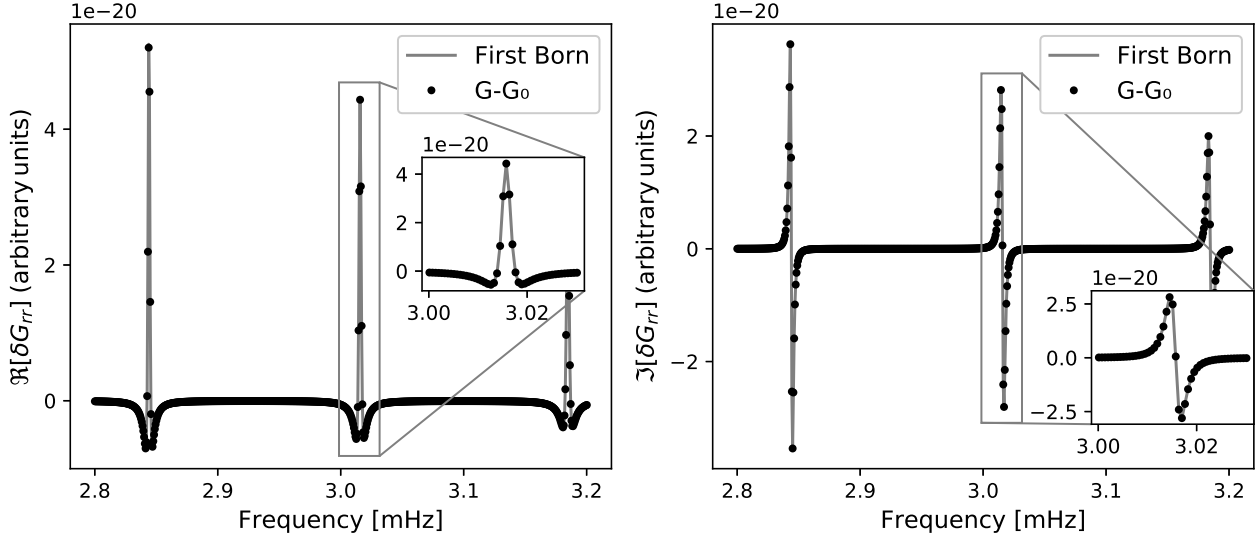}\hfill{}

\caption{\label{fig:dG}Change in the radial component of the Green function
($\delta G_{rr}$) for an isotropic perturbation in sound speed, computed
using the first Born approximation (line) and as the difference between
the Green functions in the two models (dots). The left panel shows
the real part of $\delta G_{rr}$ and the right one shows the imaginary
component. The Green function has been computed for $\ell=40$, and only a section of the frequency range has been displayed.}

\end{figure*}

We verify Equation \eqref{eq:deltaG_int_angle} by comparing the radial
component of $\delta\mathbf{G}$ for an isotropic sound-speed perturbation
$\delta c\left(r\right)$ with that computed from the difference in
the full Green functions. A spherically symmetric perturbation $\delta c\left(r\right)$
may be expanded in a basis of spherical harmonics in terms of only
the component corresponding to $\ell=0$ and $m=0$ as $\delta c\left(r\right)=\delta c_{00}\left(r\right)/\sqrt{4\pi}$.
The imposition of $\ell=0$ necessitates $j_{2}=j_{1}$, and $m=0$
implies $m_{1}=-m_{2}$. Limiting the sum in Equation \eqref{eq:deltaG_int_angle}
to only the non-zero terms, we obtain the expression for the radial
component to be
\begin{align}
\delta G_{rr}\left(\mathbf{x}_{\text{obs}},\mathbf{x}_{\text{src}},\omega\right) =& -2\int r^{2}dr\,\rho\, c\,\delta c\,\left(r\right) \times \notag \\ & \sum_{j}\frac{\left(2j+1\right)}{4\pi} \left[\grad\cdot\mathbf{G}\right]_{0j\omega}\left(r,r_{\text{obs}}\right)\times \notag \\ & \left[\grad\cdot\mathbf{G}\right]_{0j\omega}\left(r,r_{\text{src}}\right)P_{j}\left(\hat{n}_{\text{obs}}\cdot\hat{n}_{\text{src}}\right).\label{eq:deltaG_int_angle_sym}
\end{align}
Similar
to \citet{2017ApJ...842...89M}, we choose $\delta c\left(r\right)=10^{-5}c\left(r\right)$ and compute $\delta G_{rr}\left(\mathbf{x}_{\text{obs}},\mathbf{x}_{\text{src}},\omega\right)$
from Equation \eqref{eq:deltaG_int_angle_sym}. Alternately such a
perturbation may be interpreted as a change in the radial profile
of the background sound speed, so we evaluate the Green functions
following the prescription in Section \ref{sec:Green-function} for
the original and altered sound-speed profiles. We compute $\delta G_{rr,\delta c}$
as $G_{rr,c+\delta c}\left(\mathbf{x}_{\text{obs}},\mathbf{x}_{\text{src}},\omega\right)-G_{rr,c}\left(\mathbf{x}_{\text{obs}},\mathbf{x}_{\text{src}},\omega\right)$,
where the subscript indicates the sound-speed profile that is used
in the background model. We plot both these functions in Figure \ref{fig:dG}.
The close match between the result in the VSH basis and that obtained
in the spherical polar basis serves to validate the formalism used
in this work.

\section{Cross-covariance\label{sec:Synthetic-cross-covariances}}

Waves in the Sun are generated stochastically by sources located all
over the solar surface. Therefore, to model this behaviour, we 
study the expected cross-covariance $C\left(\mathbf{x}_{1},\mathbf{x}_{2},\omega\right)$
between line-of-sight projected Doppler wave velocities measured at
points $\mathbf{x}_{1}$ and $\mathbf{x}_{2}$ on the solar surface. This quantity is obtained by ensemble averaging over many stochastic source distributions.
We denote the line-of-sight vector at the point $\mathbf{x}_{i}$
on the solar surface by $\bm{l}\left(\hat{n}_{i}\right)$. We express
the cross-covariance in temporal frequency domain as 
\begin{equation}
C\left(\mathbf{x}_{1},\mathbf{x}_{2};\omega\right)=\omega^{2}\left\langle \bm{l}\left(\mathbf{x}_{1}\right)\cdot\bm{\xi}^{*}\left(\mathbf{x}_{1};\omega\right)\bm{l}\left(\mathbf{x}_{2}\right)\cdot\bm{\xi}\left(\mathbf{x}_{2};\omega\right)\right\rangle ,\label{eq:cross_cov}
\end{equation}
where $\bm{l}\left(\mathbf{x}_{i}\right)$ is the line-of-sight vector
from the point $\mathbf{x}_{i}$ on the Sun directed towards the detector,
$\bm{\xi}\left(\mathbf{x};\omega\right)$ is the wave displacement,
and the angular brackets denote an ensemble average. Our results may
be compared to that obtained by \citet{2016ApJ...824...49B} by setting
$\bm{l}\left(\mathbf{x}_{i}\right)=\mathbf{e}_{r}$, and we refer
to this as the radial approximation henceforth. We note that the line-of-sight
projection breaks the spherical symmetry inherent to the
system, since the line-of-sight is computed with respect to a detector
fixed in space. This issue is skirted in the radial assumption, in
which case, the measured cross-covariance is independent of the detector
location and depends only on the angular distance between the observation
points.

Seismic wave displacements are related to the sources exciting them
through the Green function in Equation~\eqref{eq:xi_S}. Following
\citet{2016ApJ...824...49B}, we assume that the wave source $\mathbf{S}\left(\mathbf{x};\omega\right)$
may be represented as a realization of a stationary stochastic process
having a covariance 
\begin{equation}
\left\langle S_{i}^{*}\left(\mathbf{x}_{1};\omega\right)S_{j}\left(\mathbf{x}_{2};\omega\right)\right\rangle = P\left(\omega\right)\delta\left(\mathbf{x}_{1}-\mathbf{x}_{2}\right)\delta_{ir}\delta_{jr}\times \frac{1}{r_{\text{src}}^{2}}\delta\left(\left|\mathbf{x}_{1}\right|-r_{\text{src}}\right),
\end{equation}
where $P\left(\omega\right)$ is the power spectrum of the sources,
and $r_{\text{src}}$ is the radial coordinate at which seismic waves
are excited. This model of the source covariance is inspired by the observation that much of the non-adiabatic pressure fluctuations that excite sound waves in the Sun take place in a thin layer of about a few hundred kilometers below the photosphere, and the fluctuations occur predominantly in filamentary downdrafts \citep{1989ApJ...342L..95S, 1991LNP...388..141N}. A better model of the covariance might be obtained from numerical simulations of near-surface layers in the Sun, although we do not explore such an approach in the present work.

In this approximation, we simplify Equation \eqref{eq:cross_cov}
to obtain 
\begin{equation}
C\left(\mathbf{x}_{1},\mathbf{x}_{2};\omega\right)=\omega^{2}P\left(\omega\right)\int d\Omega_{\text{src}}\,\bm{l}\left(\mathbf{x}_{1}\right)\cdot\mathbf{G}_{r}^{*}\left(\mathbf{x}_{1},\mathbf{x}_{\text{src}};\omega\right) \bm{l}\left(\mathbf{x}_{2}\right)\cdot\mathbf{G}_{r}\left(\mathbf{x}_{2},\mathbf{x}_{\text{src}};\omega\right),\label{eq:C_G}
\end{equation}
where $\mathbf{G}_{r}\left(\mathbf{x}_{i},\mathbf{x}_{\text{src}};\omega\right)=\mathbf{G}\left(\mathbf{x}_{i},\mathbf{x}_{\text{src}};\omega\right)\cdot\hat{n}_{\text{src}}$
is the Green vector corresponding to a radial source, and the integral
is carried out over the angular distribution of sources in the Sun.
We carry out the integral to obtain 
\begin{equation}
C\left(\mathbf{x}_{1},\mathbf{x}_{2};\omega\right)=\omega^{2}P\left(\omega\right)\sum_{\alpha,\beta=-1}^{1}\sum_{jm}\gfnomega{\alpha*}0j\left(r_{1},r_{\text{src}}\right) \gfnomega{\beta}0j\left(r_{2},r_{\text{src}}\right)\bm{l}\left(\mathbf{x}_{1}\right)\cdot\vgshj{\alpha*}\left(\hat{n}_{1}\right)\bm{l}\left(\mathbf{x}_{2}\right)\cdot\vgshj{\beta}\left(\hat{n}_{2}\right).\label{eq:C_YY}
\end{equation}
We introduce the line-of-sight projected bipolar vector spherical harmonic
\begin{equation}
\sgshl{j_{1}j_{2},\alpha\beta}\left(\mathbf{x}_{\text{1}},\mathbf{x}_{2}\right) = \sum_{m_{1}m_{2}}C_{j_{1}m_{1}j_{2}m_{2}}^{\ell m} \bm{l}\left(\mathbf{x}_{\text{1}}\right)\cdot\mathbf{Y}_{j_{1}m_{1}}^{\alpha}\left(\hat{n}_{\text{1}}\right) \bm{l}\left(\mathbf{x}_{2}\right)\cdot\mathbf{Y}_{j_{2}m_{2}}^{\beta}\left(\hat{n}_{2}\right),\label{eq:Yj1j2l}
\end{equation}
and rewrite Equation \eqref{eq:C_YY} in a concise notation as
\begin{equation}
C\left(\mathbf{x}_{1},\mathbf{x}_{2};\omega\right)=\omega^{2}P\left(\omega\right)\sum_{\alpha,\beta=-1}^{1}\sum_{j}\left(-1\right)^{j}\sqrt{2j+1}  \gfnomega{\alpha*}0j\left(r_{1},r_{\text{src}}\right)\gfnomega{\beta}0j\left(r_{2},r_{\text{src}}\right)Y_{00}^{jj,\alpha\beta}\left(\mathbf{x}_{\text{1}},\mathbf{x}_{2}\right).\label{eq:C}
\end{equation}

The bipolar function $\sgshl{j_{1}j_{2},\alpha\beta}\left(\mathbf{x}_{\text{1}},\mathbf{x}_{2}\right)$ does not transform as a spherical tensor under rotation in general,
as the line-of-sight projection explicitly breaks the spherical symmetry.
In the radial assumption, we set $\bm{l}\left(\mathbf{x}_{i}\right)=\mathbf{e}_{r}$
and obtain $\sgshl{j_{1}j_{2},\alpha\beta}\left(\mathbf{x}_{\text{1}},\mathbf{x}_{2}\right)=\sgshl{j_{1}j_{2},00}\left(\mathbf{x}_{\text{1}},\mathbf{x}_{2}\right)=Y_{j_{1}j_{2}}^{\ell m}\left(\hat{n}_{\text{1}},\hat{n}_{2}\right)$,
which is the ordinary bipolar spherical harmonic and transforms
as a spherical tensor. We have chosen the notation to denote the parallel
with bipolar spherical harmonics. In particular, the function $\sgshl{j_{1}j_{2},\alpha\beta}\left(\mathbf{x}_{\text{1}},\mathbf{x}_{2}\right)$
may be written as 
\[
\sgshl{j_{1}j_{2},\alpha\beta}\left(\mathbf{x}_{\text{1}},\mathbf{x}_{2}\right)=\left[\bm{l}\left(\mathbf{x}_{\text{1}}\right)\bm{l}\left(\mathbf{x}_{2}\right)\right]:\vgshl{j_{1}j_{2},\alpha\beta}\left(\hat{n}_{1},\hat{n}_{2}\right),
\]
where the colon denotes a double contraction $\left(\mathbf{A}:\mathbf{B}=A_{ij}\,B_{ij}\right)$, and $\vgshl{j_{1}j_{2},\alpha\beta}\left(\hat{n}_{1},\hat{n}_{2}\right)$ is a bipolar vector spherical harmonic. This allows us to separate out the transformation under rotation into two parts, one for the bipolar harmonic and one for the line-of-sight tensor. The rotation matrix for the former is the Wigner D-matrix, whereas the latter may be evaluated by explicitly evaluating the line-of-sight vectors at the two points. The analysis is simplified in the radial approximation where the line-of-sight tensor $\mathbf{e}_r\mathbf{e}_r$ does not change its form on rotations.

As a sanity check, we look at the scenario where the line-of-sight
is assumed to be radial at both the observation points, which, in our
notation, implies retaining only the term corresponding to $\alpha=\beta=0$
in Equation \eqref{eq:C}. The angular term $Y_{00}^{jj,00}\left(\mathbf{x}_{\text{1}},\mathbf{x}_{2}\right)$
reduces to the bipolar spherical harmonic $Y_{00}^{jj}\left(\hat{n}_{1},\hat{n}_{2}\right)$
in this approximation. We use $Y_{00}^{jj}\left(\hat{n}_{\text{1}},\hat{n}_{2}\right)=\left(-1\right)^{j}\sqrt{2j+1}P_{j}\left(\hat{n}_{\text{1}}\cdot\hat{n}_{2}\right)/4\pi$
--- where $P_{j}$ is the Legendre polynomial of degree $j$ ---
to obtain
\begin{equation}
C_{r}\left(\mathbf{x}_{1},\mathbf{x}_{2};\omega\right)=\omega^{2}P\left(\omega\right)\sum_{j}\frac{\left(2j+1\right)}{4\pi}\gfnomega{0*}0j\left(r_{1},r_{\text{src}}\right) \gfnomega 00j\left(r_{2},r_{\text{src}}\right)P_{j}\left(\hat{n}_{\text{1}}\cdot\hat{n}_{2}\right).\label{eq:C_radial}
\end{equation}

In the PB VSH basis, $\gfnomega 00j=G_{rr,j\omega}$.
The expression in Equation \eqref{eq:C_radial} is precisely the result
that we expect if we choose to compute the cross-covariance between
the radial components of the displacement. The other combinations
of $\alpha$ and $\beta$ in Equation \eqref{eq:C} arise as projections
of the tangential wave components in the direction of the line-of-sight.

\begin{figure*}
\includegraphics[scale=0.55]{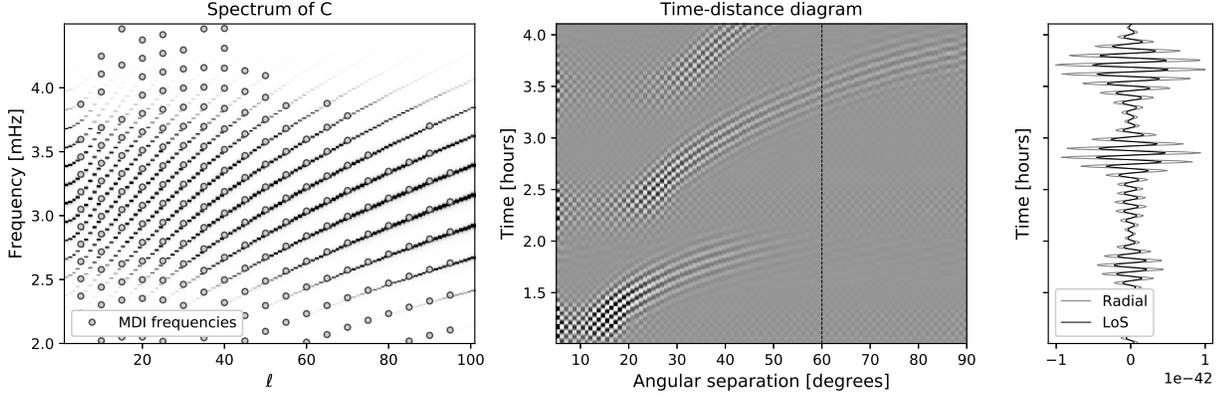}

\caption{\label{fig:C_los}Left panel: Spectrum of the cross-covariance of
radial components of  wave displacement, overlain with selected
mode frequencies obtained from the MDI pipeline \citep{1999ApJ...523L.181S}.
Middle panel: Time-distance diagram obtained for the radial component
of the displacement. Right panel: The cross-covariance between two
points on the equator at $\phi=0$ and $\phi=60$, computed using
the radial components of the displacement (grey) and the line-of-sight
projected component of the displacement (black).}
\end{figure*}

We plot the time-domain cross-covariances in Figure \ref{fig:C_los}
for one observation point on the equator at $\phi=0$ at a height
of $200\,\text{km}$ above the photosphere, and the second point at
various different azimuths on the equator and at the same observation
height, assuming a Gaussian $P\left(\omega\right)$ with a mean of
$2\pi\times3\,\text{mHz}$ and a width of $2\pi\times0.4\,\text{mHz}$.
We choose this functional form for the temporal spectrum $P\left(\omega\right)$
in numerical evaluations subsequently in the analysis.

\subsection{Change in cross-covariance}

A local change in the background sound speed by $\delta c\left(\mathbf{x}\right)$
would alter the Green function by an amount $\delta\mathbf{G}$ that
is described by Equation \eqref{eq:deltaG_int_angle}. The change
in the Green function will lead to a corresponding variation in the
measured wave fields at the two observation points, and subsequently
to a difference in the cross covariance. We may express this change as
\begin{align}
\delta C\left(\mathbf{x}_{1},\mathbf{x}_{2};\omega\right) =\omega^{2}P\left(\omega\right)\int d\Omega_{\text{src}}\,
\left[ \bm{l}\left(\mathbf{x}_{1}\right)\cdot\delta\mathbf{G}_{r}^{*}\left(\mathbf{x}_{1},\mathbf{x}_{\text{src}};\omega\right) \bm{l}\left(\mathbf{x}_{2}\right)\cdot\mathbf{G}_{r}\left(\mathbf{x}_{2},\mathbf{x}_{\text{src}};\omega\right) \right. \notag \\
 \left. + \bm{l}\left(\mathbf{x}_{1}\right)\cdot\mathbf{G}_{r}^{*}\left(\mathbf{x}_{1},\mathbf{x}_{\text{src}};\omega\right)  
  \bm{l}\left(\mathbf{x}_{2}\right)\cdot\delta\mathbf{G}_{r}\left(\mathbf{x}_{2},\mathbf{x}_{\text{src}};\omega\right)\right],\label{eq:dC_G_3D}
\end{align}

where $\delta\mathbf{G}$ is given by Equation \eqref{eq:deltaG_int_angle}.
We compute the change in the cross covariance in terms of the spherical-harmonic coefficients of the sound-speed perturbation, 
\begin{align}
\delta C\left(\mathbf{x}_{1},\mathbf{x}_{2};\omega\right) =& \omega^{2}P\left(\omega\right)\sum_{\ell m}\int r^{2}dr\delta c_{\ell m}\left(r\right)\sum_{\delta\left(j_{1}j_{2},\ell\right)} N_{j_{1}j_{2}\ell} \notag \\ 
& \sum_{\alpha\beta} T_{\alpha,j_{1}j_{2}\omega}^{\beta}\left(r,r_{1},r_{2},r_{\text{src}}\right)\sgshl{j_{1}j_{2},\alpha\beta}\left(\mathbf{x}_{1},\mathbf{x}_{2}\right),\label{eq:deltaC}
\end{align}
where the terms $N_{j_{1}j_{2}\ell}$ and $T_{\alpha,j_{1}j_{2}\omega}^{\beta}$
are defined as 
\begin{align}
N_{j_{1}j_{2}\ell}  =& \sqrt{\frac{\left(2j_{1}+1\right)\left(2j_{2}+1\right)}{4\pi\left(2\ell+1\right)}}C_{j_{1}0j_{2}0}^{\ell0},\label{eq:Nj1j2l}\\
T_{\alpha,j_{1}j_{2}\omega}^{\beta}\left(r,r_{1},r_{2},r_{\text{src}}\right) =& H_{\alpha,j_{1}j_{2}\omega}^{\beta}\left(r,r_{1},r_{2},r_{\text{src}}\right)+ \notag \\ & H_{\beta,j_{2}j_{1}\omega}^{\alpha*}\left(r,r_{2},r_{1},r_{\text{src}}\right),\\
H_{\alpha,j_{1}j_{2}\omega}^{\beta}\left(r,r_{1},r_{2},r_{\text{src}}\right) =& -2\rho c\left[\grad\cdot\mathbf{G}\right]_{\alpha,j_{1}\omega}^{*}\left(r,r_{1}\right) \times \notag \\ & \left[\grad\cdot\mathbf{G}\right]_{0,j_{2}\omega}^{*}\left(r,r_{\text{src}}\right) \times \notag \\ & \gfnomega{\beta}0{j_{2}}\left(r_{2},r_{\text{src}}\right),
\end{align}
and $\delta\left(j_{1}j_{2},\ell\right)$ as a summation index indicates
a sum over all whole-number values of $j_{1}$ and $j_{2}$ that are
related to $\ell$ through the triangle inequality $\left|j_{1}-\ell\right|\leq j_{2}\leq j_{1}+\ell$.
We note that $C_{j_{1}0j_{2}0}^{\ell0}$ is non-zero only if $j_{1}+j_{2}+\ell$
is even, which implies that contributions towards sound-speed modes
with odd-$\ell$ come only from wave-mode pairs $\left(j_{1},j_{2}\right)$
for which $j_{1}+j_{2}$ is odd, and similarly for even ones. This
limits the number of terms that contribute to the summation in Equation
\eqref{eq:deltaC}.

An isotropic sound-speed perturbation $\delta c\left(r\right)$ would
lead to the angular variation being described by $Y_{00}^{j_{1}j_{2},\alpha\beta}\left(\mathbf{x}_{\text{1}},\mathbf{x}_{2}\right).$
We recognize that this term has the same angular variation as the cross-covariance
from Equation \eqref{eq:C}. This is consistent with the idea that
a radial perturbation $\delta c\left(r\right)$ changes the background
sound speed from $c\left(r\right)$ to $c\left(r\right)+\delta c\left(r\right)$
while still preserving spherical symmetry of the medium, and we may
therefore evaluate the change in cross covariance directly as the
difference in the cross covariances computed in each model.

\subsection{Numerical evaluation of change in cross-covariance}

We use the publicly available library SHTOOLS \citep{2018GGG....19.2574W}
to compute the Clebsch-Gordan coefficients that enter Equations \eqref{eq:Yj1j2l}
and \eqref{eq:Nj1j2l}. The library can produce reliable estimates
of Clebsch Gordan coefficients till an angular degree of around $160$.
We compute the PB VSH using the fact that the generalized spherical
harmonics $Y_{jm}^{\alpha}$ may be computed as phase-shifted elements
of the Wigner d-matrix $d_{-\alpha-m}^{j}$. We compute the Wigner
d-matrix elements through an exact diagonalization of the angular
momentum operator $J_{y}$ following the prescription laid out by
\citet{PhysRevE.92.043307}. This produces Wigner d-matrices with
an accuracy of $10^{-14}$ for $j\leq100$. We therefore limit ourselves
to $j_{1},j_{2}\leq80$ in this work, so that by the triangle inequality
we obtain $\ell\leq j_{1}+j_{2}\leq160$. Assuming a
radial line-of-sight reduces the generalized spherical harmonics $Y_{jm}^{\alpha}$
to regular spherical harmonics $Y_{jm}$ which may be computed accurately
till much higher orders, for example with absolute or relative errors
$\leq10^{-10}$ up to $j\sim1000$ using the recursive algorithm prescribed
by \citet{Limpanuparb2014arXiv1410.1748L}.

We verify our result by using an isotropic sound-speed perturbation
$\delta c\left(r\right)=\delta c_{00}\left(r\right)/\sqrt{4\pi}$.
An isotropic sound-speed perturbation may be interpreted as a change
in the background sound speed profile to $c\left(r\right)+\delta c\left(r\right)$,
and we may compute the cross covariance in this model using Equation
\eqref{eq:C_G}. We choose a specific model of the perturbation given
by $\delta c\left(r\right)=10^{-5}c\left(r\right)$. We compute $\delta C_{\delta c}\left(\mathbf{x}_{1},\mathbf{x}_{2};\omega\right)=C_{c+\delta c}\left(\mathbf{x}_{1},\mathbf{x}_{2};\omega\right)-C_{c}\left(\mathbf{x}_{1},\mathbf{x}_{2};\omega\right)$
--- where the subscript indicates the sound-speed profile in the
background model --- and compare it with 
\begin{widetext}
\begin{align}
\delta C\left(\mathbf{x}_{1},\mathbf{x}_{2};\omega\right)\, =\, & \omega^{2}P\left(\omega\right)\int r^{2}dr\delta c\left(r\right) \sum_{j}\left(-1\right)^{j}\sqrt{2j+1} \sum_{\alpha,\beta=-1}^{1}T_{\alpha,jj\omega}^{\beta}\left(r,r_{1},r_{2},r_{\text{src}}\right)Y_{00}^{jj,\alpha\beta}\left(\mathbf{x}_{1},\mathbf{x}_{2}\right).
\end{align}
\end{widetext}

\begin{figure*}
\includegraphics[scale=0.8]{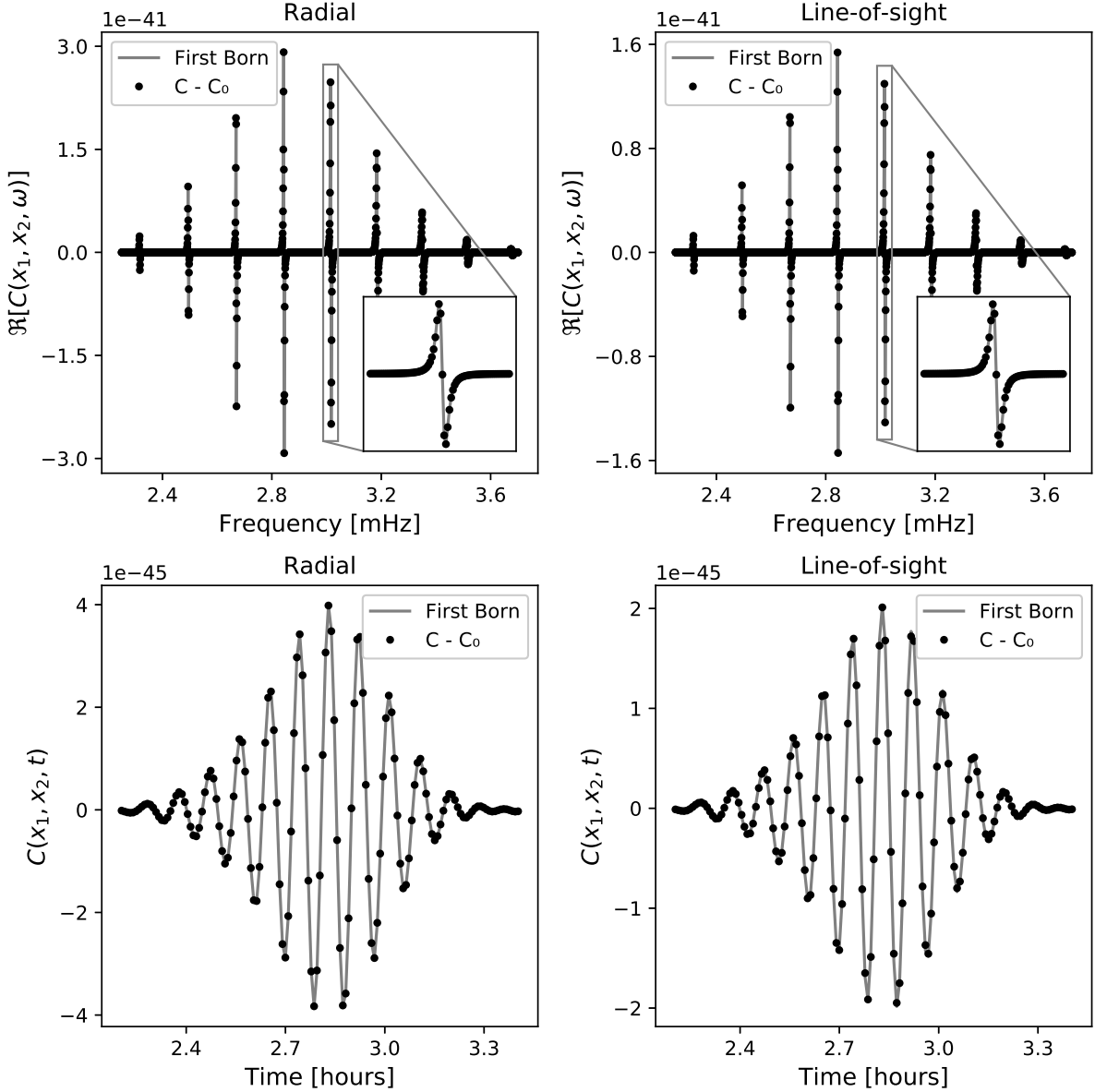}

\caption{\label{fig:deltaC}Change in cross-covariance computed using the first
Born approximation (solid gray line) and as the direct difference
between the cross covariances computed using two different models
of radial sound-speed profiles. The top row represents $\delta C$ in frequency domain for $\ell=40$, whereas the bottom row represents the $\delta C$ in time domain, computed by summing over all modes lying in $5\leq\ell\leq80$. The left panel is computed using the
radial components of wave displacement, whereas the right panel is
computed using the line-of-sight projected components.}
\end{figure*}

We plot the two functions in Figure \ref{fig:deltaC}. We also compare the change in cross-covariances using only the radial components. We find that the significant difference introduced by the projection is in the amplitude of the change in cross-covariance.

\section{Sensitivity kernel\label{sec:kernel}}

\subsection{Travel time}

A change in propagation speeds would leave its imprint on the time
taken by waves to travel between two points on the Sun. An estimate
of this change may be determined by following \citet{2002ApJ...571..966G}
as 
\begin{equation}
\delta\tau\left(\mathbf{x}_{1},\mathbf{x}_{2}\right)=\int_{0}^{\infty}\frac{d\omega}{2\pi}\,2\Re\left[h^{*}\left(\mathbf{x}_{1},\mathbf{x}_{2},\omega\right)\delta C\left(\mathbf{x}_{1},\mathbf{x}_{2},\omega\right)\right],\label{eq:dtau_dC}
\end{equation}
where $h\left(\mathbf{x}_{1},\mathbf{x}_{2},\omega\right)$ is a weighing
function that depends on the specific measurement technique and is
computed in terms of the cross-covariance $C\left(\mathbf{x}_{1},\mathbf{x}_{2},\omega\right)$.
We may use Equation \eqref{eq:deltaC} to recast Equation \eqref{eq:dtau_dC}
in the form

\begin{equation}
\delta\tau\left(\mathbf{x}_{1},\mathbf{x}_{2}\right) =\sum_{\ell m}\int_{0}^{R\odot}dr\,r^{2}K_{\ell m}^{*}\left(r,\mathbf{x}_{1},\mathbf{x}_{2}\right)\delta c_{\ell m}\left(r\right),\label{eq:inv_prob_c}
\end{equation}
where the kernel components $K_{\ell m}\left(r,\mathbf{x}_{1},\mathbf{x}_{2}\right)$
are given by 

\begin{equation}
K_{\ell m}\left(r,\mathbf{x}_{1},\mathbf{x}_{2}\right) = \int_{0}^{\infty}\frac{d\omega}{2\pi}\,\omega^{2}P\left(\omega\right)\, \sum_{\delta\left(j_{1}j_{2},\ell\right)}\sum_{\alpha\beta}N_{j_{1}j_{2}\ell}\,
2\Re\left[h^{*}\left(\mathbf{x}_{1},\mathbf{x}_{2},\omega\right)T_{\alpha,j_{1}j_{2}\omega}^{\beta}\left(r,r_{1},r_{2},r_{\text{src}}\right)\right]
\sgshl{j_{1}j_{2},\alpha\beta*}\left(\mathbf{x}_{1},\mathbf{x}_{2}\right).\label{eq:Klm}
\end{equation}

\begin{figure*}
\includegraphics[scale=0.8]{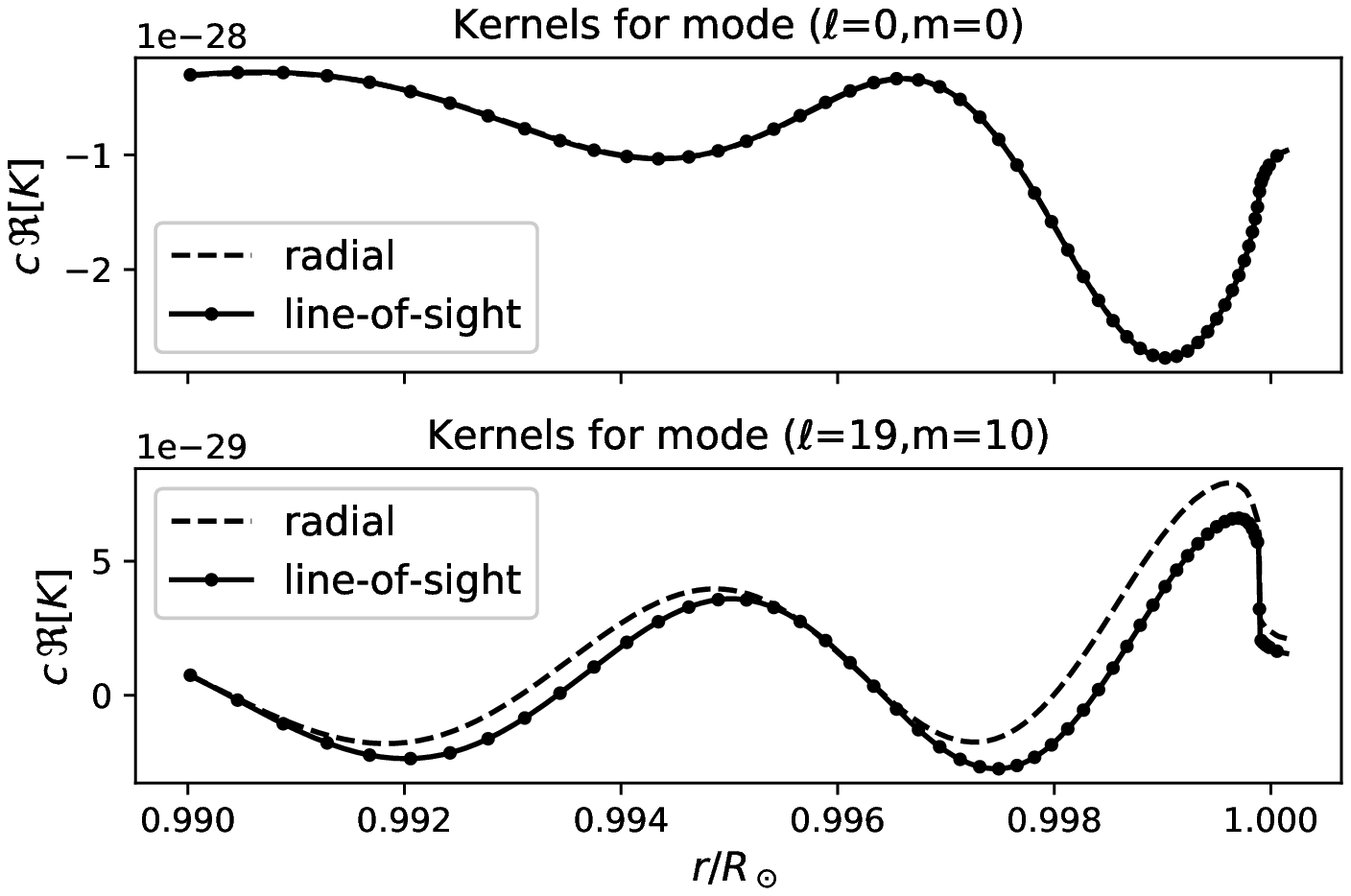}

\caption{\label{fig:kernel_profiles}Profiles of kernel components for modes $(\ell=0,m=0)$ (top) and $(\ell=19,m=10)$ (bottom), computed using radial and line-of-sight projected wave displacements.}
\end{figure*}

Equation \eqref{eq:inv_prob_c} sets up an inverse problem for the
sound-speed perturbation in terms of the measured travel times $\delta\tau\left(\mathbf{x}_{1},\mathbf{x}_{2}\right)$,
where the kernel incorporates the effect of line-of-sight projection
and variation in the center-to-limb observational heights. We plot
the radial profile of the kernel for various $\left(\ell,m\right)$
pairs in Figure \ref{fig:kernel_profiles} in the radial line-of-sight
approximation, choosing two observation points located at a co-latitude of $45$ degrees, and having azimuths of $-45$ degrees and $45$ degrees respectively. We need only compute $K_{\ell m}$ for
$m\geq0$ in the inverse problem in Equation \eqref{eq:inv_prob_c},
as $\delta c\left(\mathbf{x}\right)$ is a real function in space.

 We may construct the three-dimensional kernel $K\left(\mathbf{x},\mathbf{x}_{1},\mathbf{x}_{2}\right)$
by summing over spherical harmonics as 
\begin{equation}
K\left(\mathbf{x},\mathbf{x}_{1},\mathbf{x}_{2}\right)=\sum_{\ell m}K_{\ell m}\left(r,\mathbf{x}_{1},\mathbf{x}_{2}\right)\shl\left(\theta,\phi\right).\label{eq:K3D_Mandal}
\end{equation}
The inverse problem in Equation \eqref{eq:inv_prob_c}
is $1.5$-dimensional, and does not require a computation of the three-dimensional
kernel.

\subsection{Validating the travel-time kernel: radial displacements}

We verify that the analysis produces the expected result by matching
the expression for the kernel in Equation \eqref{eq:Klm} with that
obtained by carrying out a spherical harmonic decomposition of the
three-dimensional fleshed-out kernel. We start by computing the cross-covariance
of the radial component of the displacement measured at two points.
The corresponding three-dimensional travel-time kernel evaluates to
\begin{align}
K\left(\mathbf{x},\mathbf{x}_{1},\mathbf{x}_{2}\right) = & \int_{0}^{\infty}\frac{d\omega}{2\pi}\,\omega^{2}P\left(\omega\right)
\sum_{j_{1}j_{2}} \frac{\left(2j_{1}+1\right)}{4\pi}\frac{\left(2j_{2}+1\right)}{4\pi}P_{j_{1}}\left(\hat{n}\cdot\hat{n}_{1}\right)P_{j_{2}}\left(\hat{n}\cdot\hat{n}_{2}\right)\times  \notag \\
& 2\Re\left[h^{*}\left(\mathbf{x}_{1},\mathbf{x}_{2},\omega\right)T_{0,j_{1}j_{2}\omega}^{0}\left(r,r_{1},r_{2},r_{\text{src}}\right)\right]
. \label{eq:kernel_3D}
\end{align}
We see that the angular dependence is given by the product of two
Legendre polynomials. Using the decomposition of Legendre polynomials
in terms of spherical harmonics, we obtain the result 

\begin{equation}
\frac{\left(2j_{1}+1\right)}{4\pi}\frac{\left(2j_{2}+1\right)}{4\pi}P_{j_{1}}\left(\hat{n}\cdot\hat{n}_{1}\right)P_{j_{2}}\left(\hat{n}\cdot\hat{n}_{2}\right) = \sum_{\ell m}N_{j_{1}j_{2}\ell}\sgshl{j_{1}j_{2}*}\left(\hat{n}_{1},\hat{n}_{2}\right)Y_{\ell m}\left(\hat{n}\right) ,
\end{equation}
where we have used the angular momentum coupling relation of spherical
harmonics \citep[see][]{1988qtam.book.....V}. We have detailed the
steps in arriving at these results in Appendix \ref{sec:3D-kernel}.
Substituting this into Equation \eqref{eq:kernel_3D}, we obtain 
\begin{align}
K_{\ell m}\left(\mathbf{x},\mathbf{x}_{1},\mathbf{x}_{2}\right) =& \int_{0}^{\infty}\frac{d\omega}{2\pi}\,\omega^{2}P\left(\omega\right)\sum_{\delta\left(j_{1}j_{2},\ell\right)}\sum_{\ell m}N_{j_{1}j_{2}\ell}\times \notag \\ 
& 2\Re\left[h^{*}\left(\mathbf{x}_{1},\mathbf{x}_{2},\omega\right)T_{0,j_{1}j_{2}\omega}^{0}\left(r,r_{1},r_{2},r_{\text{src}}\right)\right] \notag \\ 
& \times \sgshl{j_{1}j_{2}*}\left(\hat{n}_{1},\hat{n}_{2}\right).
\end{align}
This matches with the expression for the spherical harmonic components
of the kernel from Equation \eqref{eq:Klm} in the radial approximation
$\left(\alpha=\beta=0\right)$, and validates the analysis. 

\begin{figure*}
\includegraphics[width=\linewidth]{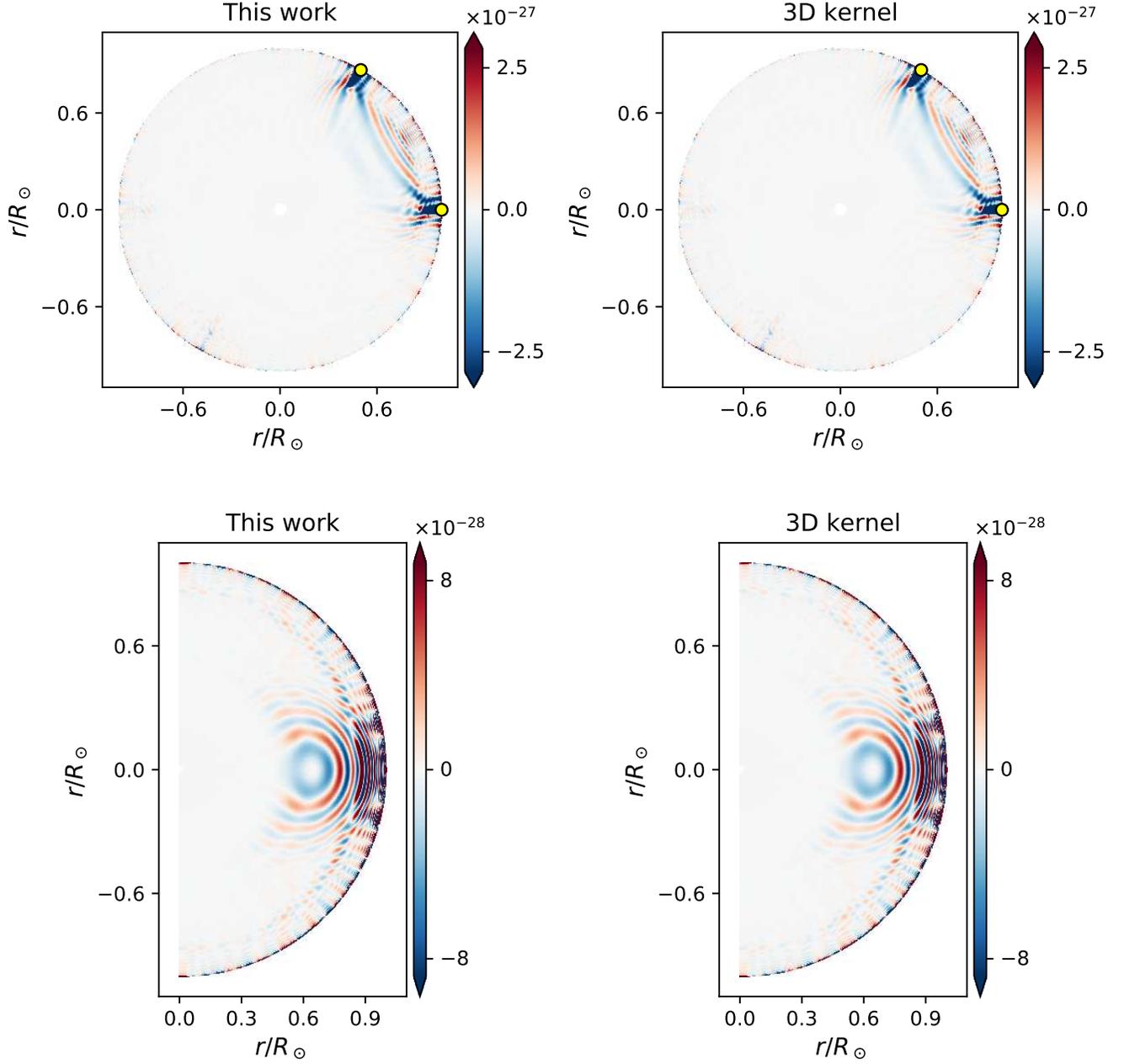}
\caption{\label{fig:K3D_Mandal_cross_section}Equatorial slice (top) and meridional slice (bottom) of $cK_{c}\left(\mathbf{x},\mathbf{x}_{1},\mathbf{x}_{2}\right)$
for $\mathbf{x}_{1}=\left(R_{\odot}+200\text{km},\pi/2,0\right)$
and $\mathbf{x}_{2}=\left(R_{\odot}+200\text{km},\pi/2,\pi/3\right)$,
computed in the radial line-of-sight assumption. The left panel in each case is
computed by summing over the spherical harmonic components using Equation
\eqref{eq:K3D_Mandal}, whereas the right panel is computed by starting
from the fleshed-out Green functions using Equation \eqref{eq:kernel_3D}.
The yellow dots denote the location of the two observation points. The meridional cross-section has been computed at $\phi=\pi/6$, which is halfway between the longitudes of the two observation points. The color scales have been saturated to highlight the deep-lying features.}
\end{figure*}

We plot an equatorial cross-section of the three-dimensional kernel
evaluated at $r=R_{\odot}$ in the top left panel of Figure \ref{fig:K3D_Mandal_cross_section},
choosing the two observation points $\mathbf{x}_{1}$ and $\mathbf{x}_{2}$
to be $\left(R_{\odot}+200\text{km},\frac{\pi}{2},0\right)$ and $\left(R_{\odot}+200\text{km},\frac{\pi}{2},\frac{\pi}{3}\right)$
respectively, and compare it with the same computed using Equation
\ref{eq:kernel_3D} in the top right panel. We impose the cutoff $j_{1},j_{2}<=80$
on the angular degrees of the waves keeping numerical accuracy in
mind, which translates to a cutoff of $\ell<=160$ on the kernel components
through the triangle inequality. We have multiplied the kernel by
the sound-speed profile to highlight the radial profile. We find a
good match between the two functions, which affirms the correctness
of the analysis. The bottom left and bottom right panels of Figure \ref{fig:K3D_Mandal_cross_section} show meridional cross-sections of the same functions evaluated at $\phi=\pi/6$.

\subsection{Validating the travel-time kernel: isotropic perturbation}

We verify the expression for the kernel by using an isotropic $\delta c\left(r\right)$
and computing wave travel-time shifts between the point $\mathbf{x}_{1}=\left(R_{\odot}+200\text{km},\frac{\pi}{2},0\right)$
and $\mathbf{x}_{2}=\left(R_{\odot}+200\text{km},\frac{\pi}{2},\phi\right)$
for various choices of $\phi$. The corresponding kernel may be obtained
by substituting $\ell=m=0$ in Equation \eqref{eq:Klm}. We also compute
the kernel for the radial components of the displacement, and compute
the travel times with the resultant expression to compare with \citet{2017ApJ...842...89M}.
The expression for the kernel using just the radial components simplifies
to 
\begin{widetext}
\begin{align}
K_{00}\left(r,\mathbf{x}_{1},\mathbf{x}_{2}\right) & =\frac{1}{\sqrt{4\pi}}\int_{0}^{\infty}\frac{d\omega}{2\pi}\,\omega^{2}P\left(\omega\right)\sum_{j}\frac{\left(2j+1\right)}{4\pi}\,2\Re\left[h^{*}\left(\mathbf{x}_{\text{1}},\mathbf{x}_{2},\omega\right)T_{0,jj\omega}^{0}\left(r,r_{1},r_{2},r_{\text{src}}\right)\right]P_{j}\left(\hat{n}_{\text{1}}\cdot\hat{n}_{2}\right),
\end{align}
\end{widetext}
where $P_{j}$ represents the Legendre polynomial of degree $j$.

\begin{figure*}
\includegraphics[scale=0.8]{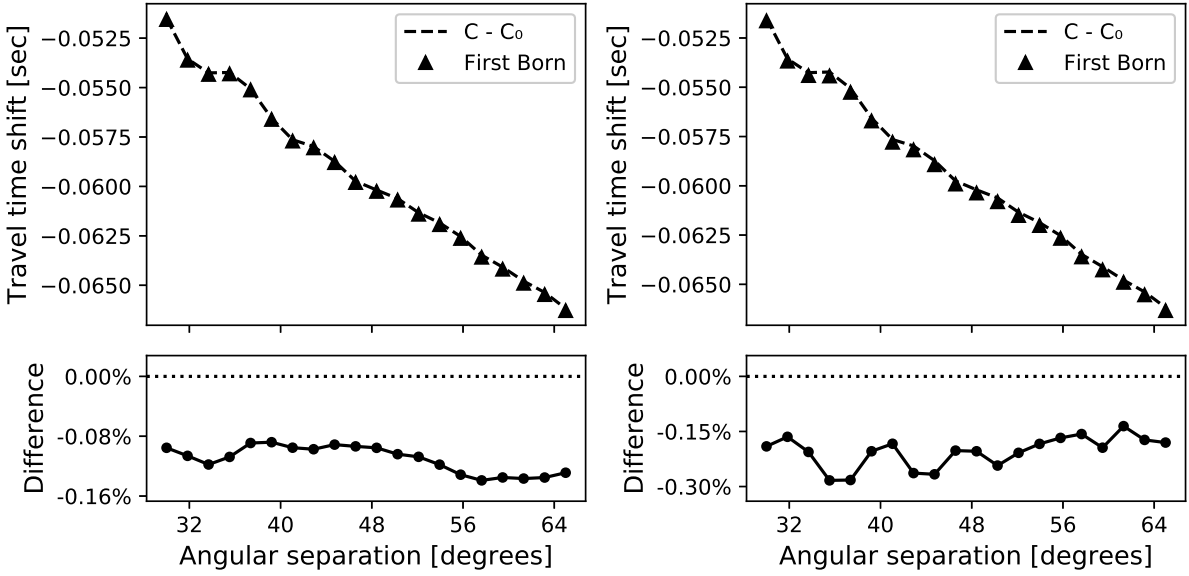}

\caption{\label{fig:sound-speed-kernel-comp}Travel time shifts between two
models having radial sound-speed profiles $c\left(r\right)\times\left(1+10^{-5}\right)$
and $c\left(r\right)$, using waves measured at the two points $\mathbf{x}_{1}=\left(R_{\odot}+200\text{km},\pi/2,0\right)$
and $\mathbf{x}_{2}=\left(R_{\odot}+200\text{km},\pi/2,\phi\right)$
for various different values of the angular separation $\phi$. The
left panel is using radial components of the waves, whereas the right
panel uses the line-of-sight projected components.}
\end{figure*}

We also compute travel time shifts using Equation \eqref{eq:dtau_dC}
for the same sets of points, where we compute the change in cross
covariance $\delta C$ as an explicit difference between the cross
covariance $C$ evaluated in the two background models. We plot both
sets of travel times in Figure \ref{fig:sound-speed-kernel-comp},
with and without line-of-sight projection. We find that the measured travel times in both the cases are quite similar. This is a consequence of the result that the travel-time kernel components for $\ell=0$ and $m=0$ are nearly identical with and without projection.

\subsection{Computational efficiency}

\begin{figure*}
\includegraphics[scale=0.74]{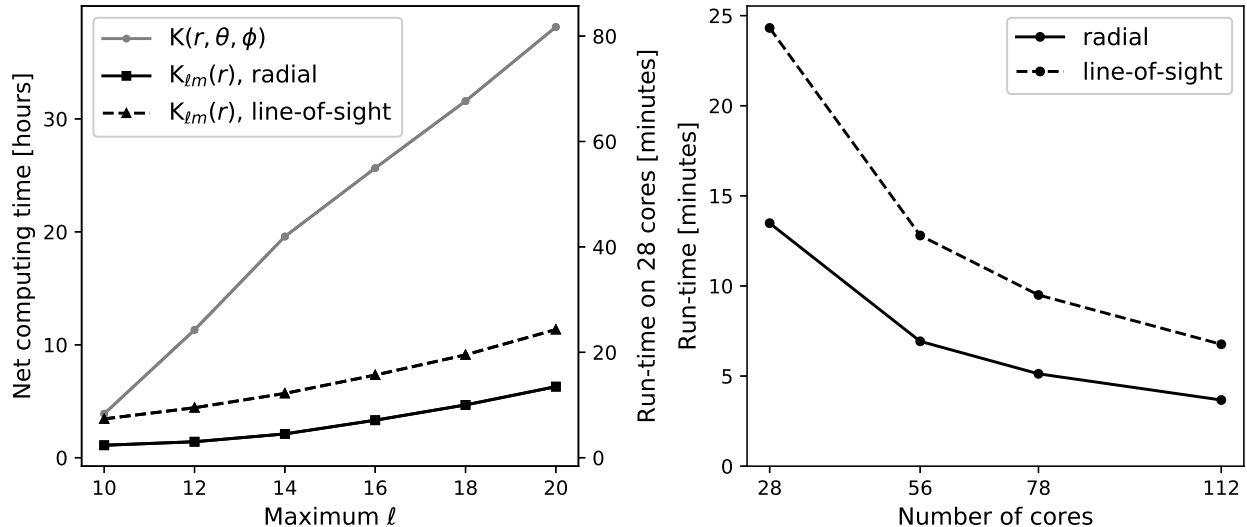}
\caption{\label{fig:runtimes}Left: Computational time required to evaluate the kernel components compared with that required to compute the three-dimensional kernel $K(r,\theta,\phi)$ following \citet{2017ApJ...842...89M}. The solid black line represents time taken to compute kernels using only the radial components of the wave displacement, whereas the dashed black line represents that for line-of-sight projected components of wave displacement. Right: Evaluation time as a function of the number of cores used compute the kernel components.}

\end{figure*}

One of the advantages of this approach over evaluating the spherical harmonic coefficients of the three-dimensional kernel $K(\mathbf{x},\mathbf{x}_2,\mathbf{x}_2)$ is that the evaluations of the kernel components are much more efficient if we are interested in a limited number of them. To demonstrate this, we evaluate the kernels $K_{\ell m}(r)$ using our approach, and the coefficients 
\begin{equation}
K^{3D}_{\ell m}(r)=\int d\Omega\, Y_{\ell m}^*(\hat{n}) K(\mathbf{x}=(r,\hat{n}),\mathbf{x}_2,\mathbf{x}_2),
\end{equation}
and compare the evaluation time in the two approaches. The latter involves two steps --- evaluating the three-dimensional profile of the kernel $K(\mathbf{x},\mathbf{x}_2,\mathbf{x}_2)$, followed by a decomposition in the basis of spherical harmonics, where the second step is significantly faster than the first one. We therefore list the time taken only to evaluate the kernel $K(\mathbf{x},\mathbf{x}_2,\mathbf{x}_2)$. In the first case we compute the kernel components for all the modes $(\ell,m)$ for $-\ell\leq m \leq \ell$ and $\ell\leq \ell_\mathrm{max}$, whereas in the latter we compute the three-dimensional profile for co-latitude $\theta$ corresponding to $\left(\ell_\mathrm{max}+1\right)/2$ Gauss-Legendre nodes, and $2\ell_\text{max}$ uniformly spaced points in the azimuthal coordinate $\phi$ ranging from $0$ to $2\pi$. The computation is carried out by summing over Green functions having angular degrees in the range $5 \leq j \leq 80$ and $4000$ uniformly spaced frequencies in the range $2\,\text{mHz}$ to $4.5\,\text{mHz}$. We arbitrarily choose the two observation points to be $\mathbf{x}_1=(R_\odot+200\,\text{km},\pi/6,-\pi/4)$ and $\mathbf{x}_2=(R_\odot+200\,\text{km},-\pi/3,\pi/4)$. We carry out the computation on one node of the Dalma cluster at New York University Abu Dhabi, that has $28$ processors available. We compare the evaluation times in Figure \ref{fig:runtimes}, where we show that the spherical-harmonic components in both the radial and line-of-sight-projected approaches may be computed significantly faster than the three-dimensional profile. The evaluation times shown in Figure \ref{fig:runtimes} are not to be treated as absolute limits as further optimization is possible; however, they suffice to demonstrate the trend in the comparative analysis presented here. The evaluation times would also be lower for points having special geometrical locations --- such as poles or being located on the Equator --- in which case various spherical-harmonic symmetries might lead to cancellations in the terms being summed. Such an approach, along with rotating the kernels components on the sphere, might lead to rapid evaluation of kernel components.

The radial parts $T_{\alpha,j_{1}j_{2}\omega}^{\beta}\left(r,r_{1},r_{2},r_{\text{src}}\right)$ of the kernels $K_{\ell m}$ in Equation. \ref{eq:Klm} do not depend on $\ell$, which provides a major computational advantage as they need to be computed only once for each combination of $(j_1,j_2,\omega)$ and may be reused for each $(\ell,m)$. This result arises from the fact that the sound-speed perturbation acts as a multiplicative scalar term in Equation \ref{eq:deltaG_int_angle}, and the result is not expected to hold for other parameters such as flow velocity. We therefore expect the computation of kernels for such parameters to be more resource intensive.

\section{Conclusion}

The inference of solar subsurface features using surface measurements relies on an accurate estimate of the sensitivity kernel that relates the model parameters to the measurements. The sensitivity kernel incorporates the physics of wave propagation as well as the systematics involved in the measurement procedure. In this work we have illustrated an approach that lets us compute travel-time sensitivity kernels in the Sun while incorporating the systematic effects introduced by spherical geometry in the measurements, such as line-of-sight projection in Dopplergram measurements and center-to-limb differences in line-formation heights. Conventional helioseismic inversions have adjusted for these systematics by correcting the measured travel-times \citep{2012ApJ...749L...5Z}, but a better understanding of the underlying physics might be obtained by a first-principle approach such as the one presented in this work.

The analysis presented in this work leads to the spherical-harmonic components of the sensitivity kernels. Large-scale features in the Sun may be expressed in an angular spherical harmonic basis in terms of a limited number of low-degree components. The present analysis therefore provides us with two advantages --- firstly the computation is numerically efficient if seek only a few low-degree modes, and secondly the inverse problem is better conditioned if we limit the number of parameters being solved for. 

In this work we have computed sensitivity kernels for sound-speed perturbations which is a scalar (rank-$0$) field. The analysis is however general in scope, and can be extended to tensors of any rank by using the definition of the angular derivative of a vector spherical harmonic \citep[see][]{1988qtam.book.....V}. Of particular interest is the analysis of large-scale subsurface flows in the Sun such as meridional flows. Recently \citet{2018ApJ...863...39M} had carried out an analysis of subsurface flows on a similar line by numerically projecting the $3$-dimensional sensitivity kernel on to a sparse basis. Our analysis demonstrates how such a decomposition can be arrived at analytically, thereby increasing the computational efficiency of the procedure.

\acknowledgments
This work was supported by NYUAD Institute Grant
G1502 "NYUAD Center for Space Science".

\appendix
{}

\section{Radial components of Green functions\label{sec:Green-functions-radial}}

We choose to rewrite the wave equation in terms of the displacement
$\bm{\xi}$, the pressure perturbation $p^{\prime}$ and the density
perturbation $\rho^{\prime}$ as 
\begin{equation}
\mathcal{L}\bm{\xi}=-\rho\omega^{2}\bm{\xi}-2i\omega\gamma\bm{\xi}+\grad p^{\prime}+\rho^{\prime}g\mathbf{e}_{r}=\mathbf{S}.\label{eq:wave_eq_p_rho}
\end{equation}
The form of the wave equation naturally suggests the use of the Hansen
VSH basis. We expand the vector wave displacement and source as 
\begin{align}
\bm{\xi}\left(\mathbf{x}\right) & =\sum_{\ell m\alpha}\xi_{\ell m}^{\left(\alpha\right)}\left(r\right)\vgshl{\left(\alpha\right)},\\
\mathbf{S}\left(\mathbf{x}\right) & =\sum_{\ell m\alpha}S_{\ell m}^{\left(\alpha\right)}\left(r\right)\vgshl{\left(\alpha\right)}.
\end{align}
and the scalar pressure and density perturbations as 
\begin{align}
p^{\prime}\left(\mathbf{x}\right) & =\sum_{\ell m}p_{\ell m}^{\prime}\left(r\right)\shl\left(\hat{n}\right),\\
\rho^{\prime}\left(\mathbf{x}\right) & =\sum_{\ell m}\rho_{\ell m}^{\prime}\left(r\right)\shl\left(\hat{n}\right).
\end{align}
Substituting these into Equation \eqref{eq:wave_eq_p_rho}, we obtain
the components of the wave equation in the Hansen VSH basis: 
\begin{align}
-\rho\omega^{2}\xi_{\ell m}^{\left(-1\right)}+\partial_{r}p_{\ell m}^{\prime}+\rho_{\ell m}^{\prime}g & =S_{\ell m}^{\left(-1\right)},\nonumber \\
-\rho\omega^{2}\xi_{\ell m}^{\left(1\right)}+\frac{\sqrt{\ell\left(\ell+1\right)}}{r}p_{\ell m}^{\prime} & =S_{\ell m}^{\left(1\right)}.\label{eq:wave_eqn_components}
\end{align}
In addition to this, we use the continuity equation 
\begin{equation}
\rho^{\prime}+\grad\cdot\left(\rho_{0}\bm{\xi}\right)=0,
\end{equation}
which is expanded in terms of the components of $\bm{\xi}$ as 
\begin{equation}
\rho_{\ell m}^{\prime}=-\frac{1}{r^{2}}\partial_{r}\left(r^{2}\rho_{0}\xi_{\ell m}^{\left(-1\right)}\right)+\frac{\sqrt{\ell\left(\ell+1\right)}}{r}\rho_{0}\xi_{\ell m}^{\left(1\right)},\label{eq:cont}
\end{equation}
and the adiabatic equation of state 
\begin{equation}
\rho_{\ell m}^{\prime}=\frac{1}{c^{2}}p_{\ell m}^{\prime}+\frac{\rho_{0}N^{2}}{g_{0}}\xi_{\ell m}^{\left(-1\right)},\label{eq:adiabatic}
\end{equation}
where the Brunt-V{\"a}is{\"a}l{\"a} frequency $N$ is defined in
terms of the thermal structure parameters $\rho_{0}$, $p_{0}$, $g_{0}$
and the adiabatic index $\Gamma_{1}$ as 
\[
N^{2}=g_{0}\left(\frac{1}{\Gamma_{1}}\frac{d\ln p_{0}}{dr}-\frac{d\ln\rho_{0}}{dr}\right).
\]
We use Equations \eqref{eq:cont} and \eqref{eq:adiabatic} to eliminate
the variables $\rho_{\ell m}^{\prime}$ and $\xi_{\ell m}^{\left(1\right)}$
from Equation \eqref{eq:wave_eqn_components} and obtain equations
in the variables $\xi_{\ell m}^{\left(-1\right)}$ and $p^{\prime}$.
The reason we choose these two variables is that we enforce the Dirichlet
boundary conditions $\xi_{\ell m}^{\left(-1\right)}\left(r=0\right)=0$
and $p^{\prime}\left(r=r_{\text{out}}\right)=0$, where $r_{\text{out}}$
is the outer boundary of the computational domain. The system of equations
for $\xi_{\ell m}^{\left(-1\right)}$ and $p^{\prime}$ therefore
becomes 
\begin{equation}
\left(\begin{array}{cc}
-\rho_{0}\left(\omega^{2}-N^{2}\right) & \partial_{r}+\frac{g_{0}}{c^{2}}\\
\partial_{r}+\frac{2}{r}-\frac{g_{0}}{c^{2}} & \frac{1}{\rho_{0}c^{2}}\left(1-\frac{S_{\ell}^{2}}{\omega^{2}}\right)
\end{array}\right)\left(\begin{array}{c}
\xi_{\ell m}^{\left(-1\right)}\\
p_{\ell m}^{\prime}
\end{array}\right)=\left(\begin{array}{c}
S_{\ell m}^{\left(-1\right)}\\
-\frac{\sqrt{\ell\left(\ell+1\right)}}{\omega^{2}}\frac{1}{r\rho_{0}}S_{\ell m}^{\left(1\right)}\left(r\right)
\end{array}\right),\label{eq:disp_matrix}
\end{equation}
and we compute the horizontal displacement $\xi_{\ell m}^{\left(1\right)}\left(r\right)$
through 
\begin{equation}
\xi_{\ell m}^{\left(1\right)}\left(r\right)=\frac{\sqrt{\ell\left(\ell+1\right)}}{\rho_{0}r\omega^{2}}p_{\ell m}^{\prime}\left(r\right)-\frac{1}{\rho_{0}\omega^{2}}S_{\ell m}^{\left(1\right)}\left(r\right).\label{eq:disp_horizontal}
\end{equation}

The Green function for the operator $\mathcal{L}$ satisfies 
\begin{equation}
\mathcal{L}\mathbf{G}\left(\mathbf{x},\mathbf{x}_{\text{src}};\omega\right)=\delta\left(\mathbf{x}-\mathbf{x}_{\text{src}}\right)\mathbf{I},\label{eq:wave_eqn_gfn_delta}
\end{equation}
where $\delta\left(\mathbf{x}-\mathbf{x}_{\text{src}}\right)$ is
the Dirac delta function centered about the point $\mathbf{x}_{\text{src}}$,
$\mathbf{I}$ is the $3\times3$ identity matrix and $\omega$ represents
temporal frequency. The wave operator $\mathcal{L}$ is spherically
symmetric in the quiet Sun, and its eigenfunctions are labeled by
radial order $n$ and angular degrees $j$ and $m$. The Green function
is expanded in the Hansen VSH basis as 
\begin{equation}
\mathbf{G}\left(\mathbf{x},\mathbf{x}_{\text{src}};\omega\right)=\sum_{\alpha,\beta=-1}^{1}\sum_{\ell m}\gfnomega{\left(\alpha\right)}{\left(\beta\right)}{\ell}\left(r,r_{\text{src}}\right)\vgshl{\left(\alpha\right)}\left(\hat{n}\right)\vgshl{\left(\beta\right)\dagger}\left(\hat{n}_{\text{src}}\right)\label{eq:green_fn_vsh}
\end{equation}
Solving for the Green function is therefore equivalent to obtaining
the radial components $G_{\left(\beta\right),\ell\omega}^{\left(\alpha\right)}\left(r,r_{\text{src}}\right)$.
We note that spherical symmetry of the wave operator implies that
the components of the Green function do not depend on the azimuthal
quantum number $m$. We also note that the delta function source is
expanded as 
\[
\delta\left(\mathbf{x}-\mathbf{x}_{\text{src}}\right)\mathbf{I}=\frac{1}{r^{2}}\delta\left(r-r_{\text{src}}\right)\sum_{\alpha=-1}^{1}\sum_{\ell m}\vgshl{\left(\alpha\right)}\left(\hat{n}\right)\vgshl{\left(\alpha\right)\dagger}\left(\hat{n}_{\text{src}}\right).
\]
Substituting these into Equation \eqref{eq:disp_matrix} we obtain
\begin{equation}
\left(\begin{array}{cc}
-\rho_{0}\left(\omega^{2}-N^{2}\right) & \partial_{r}+\frac{g_{0}}{c^{2}}\\
\partial_{r}+\frac{2}{r}-\frac{g_{0}}{c^{2}} & \frac{1}{\rho_{0}c^{2}}\left(1-\frac{S_{\ell}^{2}}{\omega^{2}}\right)
\end{array}\right)\left(\begin{array}{c}
\gfnlomega{\left(-1\right)}{\left(\beta\right)}\left(r,r_{\text{src}}\right)\\
p_{\left(\beta\right),\ell\omega}^{\prime}\left(r,r_{\text{src}}\right)
\end{array}\right)=\frac{1}{r^{2}}\delta\left(r-r_{\text{src}}\right)\left(\begin{array}{c}
\delta_{\beta}^{-1}\\
-\frac{\sqrt{\ell\left(\ell+1\right)}}{\omega^{2}}\frac{1}{r\rho_{0}}\delta_{\beta}^{1}
\end{array}\right),\label{eq:Gfn_matrix}
\end{equation}
and subsequently using \eqref{eq:disp_horizontal} we obtain 
\begin{equation}
G_{\left(\beta\right),\ell\omega}^{\left(1\right)}\left(r,r_{\text{src}}\right)=\frac{\sqrt{\ell\left(\ell+1\right)}}{\rho_{0}r\omega^{2}}\left(p_{\left(\beta\right),\ell\omega}^{\prime}\left(r,r_{\text{src}}\right)-\frac{1}{r^{2}}\delta\left(r-r_{\text{src}}\right)\delta_{\beta}^{1}\right).\label{eq:Gfn_horizontal}
\end{equation}

The components of the Green function obey certain symmetry conditions
that arise from that fact that the eigenfunctions of the wave operator
$\mathcal{L}$ are strictly spheroidal, and may be expanded in the
Hansen VSH basis in terms of the vectors $\vgshl{\left(-1\right)}$
and $\vgshl{\left(1\right)}$ without any component along $\vgshl{\left(0\right)}$.
This is seen from Equation \eqref{eq:wave_eqn_components} by noting
that the restoring force has no component along $\vgshl{\left(0\right)}$.
This implies that the summations over $\alpha$ and $\beta$ in Equation
\eqref{eq:green_fn_vsh} only range over $\pm1$. This leads to the
following symmetry conditions in the PB VSH basis: 
\begin{equation}
\begin{aligned}\gfnlomega{\alpha}+ & =\gfnlomega{\alpha}-,\quad\alpha\in\left\{ +,0,-\right\} ,\\
\gfnlomega +{\alpha} & =\gfnlomega -{\alpha},\quad\alpha\in\left\{ +,0,-\right\} .
\end{aligned}
\label{eq:spheroidal_symmetry}
\end{equation}
These symmetry relations imply that there are four independent components
of the Green's function. Without loss of generality, we choose the
components 
\begin{equation}
\begin{aligned}\gfnlomega 00 & =\gfnlomega{\left(-1\right)}{\left(-1\right)},\quad\gfnlomega +0=\frac{1}{\sqrt{2}}\gfnlomega{\left(1\right)}{\left(-1\right)},\\
\gfnlomega 0+ & =\frac{1}{\sqrt{2}}\gfnlomega{\left(-1\right)}{\left(1\right)},\quad\gfnlomega ++=\frac{1}{2}\gfnlomega{\left(1\right)}{\left(1\right)}.
\end{aligned}
\label{eq:hansen_tromp_gfn_comp}
\end{equation}
These represent the radial and tangential responses to a radial and
tangential source respectively, where the tangential direction is
specifically oriented along $\vgshl +$ at each point on the Sun.
We solve Equations \eqref{eq:disp_matrix} and \eqref{eq:disp_horizontal}
for the Hansen VSH components $\gfnlomega{\left(\pm1\right)}{\left(\pm1\right)}$
of the Green function, and use these to compute the PB VSH components
$\gfnlomega{\pm1}{\pm1}$ using Equation \eqref{eq:hansen_tromp_gfn_comp}.

\begin{figure}
\includegraphics[scale=0.6]{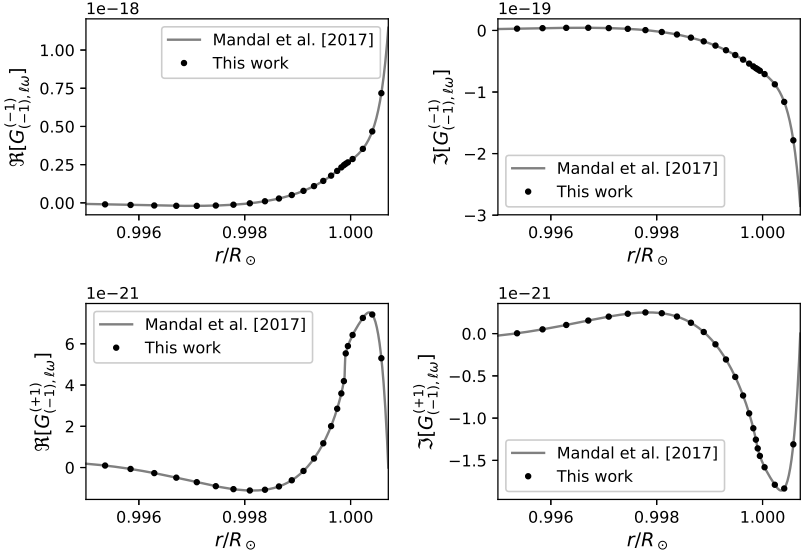}

\caption{\label{fig:Gfn_Mandal}The Green functions computed in this work compared
to those computed earlier by \citet{2017ApJ...842...89M}. The top
row shows the real and imaginary parts of the radial component of
the Green function for a radial source, whereas the bottom row shows
the tangential component of the Green function for a radial source.
The Green function has been computed for $\ell=40$ and $\nu=3\,\text{mHz}$. }

\end{figure}

Equation \eqref{eq:disp_matrix} is equivalent to that obtained by
\citet{2017ApJ...842...89M}, where the authors had solved for just
the components $\gfnlomega{\left(-1\right)}{\left(-1\right)}$ and
$\gfnlomega{\left(+1\right)}{\left(-1\right)}$ corresponding to radial
and tangential displacements for a radially directed source respectively.
The other components appear in the expression for the kernel if we
consider line-of-sight projection effects, so we choose to solve for
all of them. We solve Equation \eqref{eq:disp_matrix} numerically
using a high-order finite-difference solver similar to the one described
by \citet{2017ApJ...842...89M} using sparse-matrix representation
of the operators implemented in the Julia language \citep{julia}.
We use Model S \citep{1996Sci...272.1286C} as our background solar
model to compute the Green functions. We compare our Green functions
to those obtained by \citet{2017ApJ...842...89M} in Figure \ref{fig:Gfn_Mandal}.

We note that Equation \eqref{eq:Gfn_matrix} implies certain discontinuities
in the Green function components at the source radius. Assuming a
radial delta-function source we integrate the equations over $\left[r_{\text{src}}-\varepsilon,\,r_{\text{src}}+\varepsilon\right]$
where $\varepsilon\rightarrow0$ to obtain 
\[
p_{\left(-1\right),\ell\omega}^{\prime}\left(r_{\text{src}}+\varepsilon,r_{\text{src}}\right)-p_{\left(-1\right),\ell\omega}^{\prime}\left(r_{\text{src}}-\varepsilon,r_{\text{src}}\right)=\frac{1}{r_{\text{src}}^{2}}.
\]
On the other hand assuming a tangential delta function source we obtain
\[
\gfnlomega{\left(-1\right)}{\left(1\right)}\left(r_{\text{src}}+\varepsilon,r_{\text{src}}\right)-\gfnlomega{\left(-1\right)}{\left(1\right)}\left(r_{\text{src}}-\varepsilon,r_{\text{src}}\right)=-\frac{\sqrt{\ell\left(\ell+1\right)}}{\omega^{2}}\frac{1}{r_{\text{src}}^{3}\rho_{0}\left(r_{\text{src}}\right)}.
\]
The discontinuity in $\gfnlomega{\left(-1\right)}{\left(1\right)}$
at $r_{\text{src}}$ implies a singularity in $\gfnlomega{\left(1\right)}{\left(1\right)}$
which is related to its derivative. We also see this from Equation
\eqref{eq:Gfn_horizontal}.

\subsection{Seismic reciprocity}

\begin{figure*}
\begin{centering}
\includegraphics[scale=0.7]{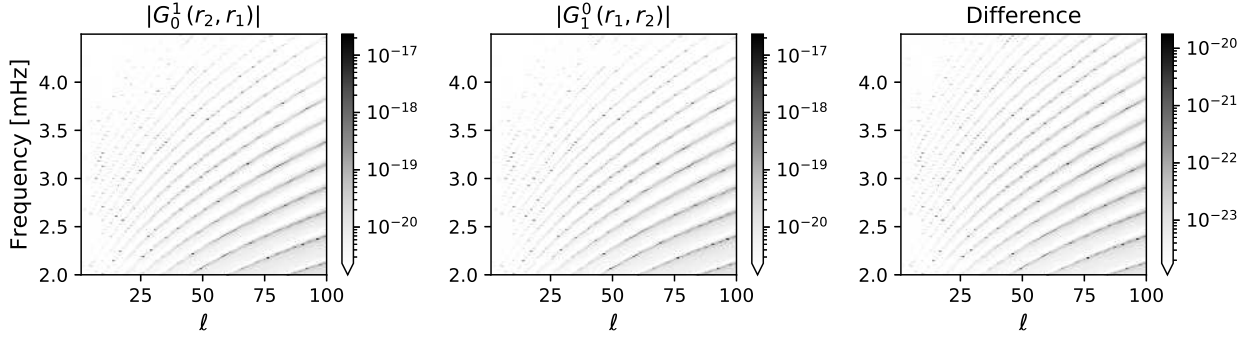}
\par\end{centering}
\caption{\label{fig:reciprocity}Absolute values of Green function components.
Left: tangential component of displacement for a radial source, middle
panel: radial component of displacement for a tangential source, right:
difference between the components.}

\end{figure*}

The wave operator $\mathcal{L}$ has an adjoint $\mathcal{L}^{\dagger}$
defined as 
\begin{equation}
\int d\mathbf{x}\,\mathbf{u}_{k}^{*}\left(\mathbf{x}\right)\cdot\mathcal{L}\mathbf{v}_{m}\left(\mathbf{x}\right)=\int d\mathbf{x}\,\left(\mathcal{L}^{\dagger}\mathbf{u}_{k}\left(\mathbf{x}\right)\right)^{*}\cdot\mathbf{v}_{m}\left(\mathbf{x}\right)\label{eq:adjoint_defn}
\end{equation}
for any pair of functions $\mathbf{u}_{k}\left(\mathbf{x}\right)$
and $\mathbf{v}_{m}\left(\mathbf{x}\right)$. The Green function $\mathbf{G}\left(\mathbf{x},\bm{\xi}\right)$
and the adjoint Green function $\mathbf{G}^{\dagger}\left(\mathbf{x},\bm{\xi}\right)$
satisfy 
\[
\mathcal{L}\mathbf{G}\left(\mathbf{x},\bm{\xi}\right)=\mathcal{L}^{\dagger}\mathbf{G}^{\dagger}\left(\mathbf{x},\bm{\xi}\right)=\delta\left(\mathbf{x}-\bm{\xi}\right)\mathbf{I}.
\]
We choose $\mathbf{u}_{k}\left(\mathbf{x}\right)$ and $\mathbf{v}_{m}\left(\mathbf{x}\right)$
in Equation \eqref{eq:adjoint_defn} to be spherical polar components
of the Green functions given by 
\begin{align*}
\mathbf{u}_{k}\left(\mathbf{x}\right) & =\mathbf{G}^{\dagger}\left(\mathbf{x},\bm{\xi}^{\prime},\omega\right)\cdot\mathbf{e}_{k}\left(\bm{\xi}^{\prime}\right),\\
\mathbf{v}_{m}\left(\mathbf{x}\right) & =\mathbf{G}\left(\mathbf{x},\bm{\xi},\omega\right)\cdot\mathbf{e}_{m}\left(\bm{\xi}\right),
\end{align*}
to obtain the relation $G_{mk}^{\dagger}\left(\bm{\xi},\bm{\xi}^{\prime}\right)=G_{km}^{*}\left(\bm{\xi}^{\prime},\bm{\xi}\right)$,
or in a compact notation $\mathbf{G^{\dagger}}\left(\bm{\xi},\bm{\xi}^{\prime}\right)=\left[\mathbf{G}\left(\bm{\xi}^{\prime},\bm{\xi}\right)\right]^{\dagger}.$
We further use the fact that $\mathcal{L}^{\dagger}=\mathcal{L}^{*}$
and hence $\mathbf{G}^{\dagger}\left(\bm{\xi},\bm{\xi}^{\prime}\right)=\mathbf{G}^{*}\left(\bm{\xi},\bm{\xi}^{\prime}\right)$
to obtain the reciprocity relation $\mathbf{G}\left(\bm{\xi}^{\prime},\bm{\xi}\right)=\mathbf{G}^{T}\left(\bm{\xi},\bm{\xi}^{\prime}\right).$
Expanding the Green tensor in the PB VSH basis we show that the components
are related by $\gfnlomega{\alpha}{\beta}\left(r,r^{\prime}\right)=\gfnlomega{-\beta}{-\alpha}\left(r^{\prime},r\right)$.
In addition to this, if we enforce the condition that the Green function
is strictly spheroidal we may use the symmetries in Equation \eqref{eq:spheroidal_symmetry}
to rewrite this as 
\begin{equation}
\gfnlomega{\alpha}{\beta}\left(r,r^{\prime}\right)=\gfnlomega{\beta}{\alpha}\left(r^{\prime},r\right).\label{eq:reciprocity}
\end{equation}
We obtain similar relations in the Hansen basis with the superscript
and subscript indices interchanged in each case. We plot $\left|\gfnlomega 10\left(r_{2},r_{1}\right)\right|$
and $\left|\gfnlomega 01\left(r_{1},r_{2}\right)\right|$ in Figure
\ref{fig:reciprocity}, choosing the values $r_{1}=R_{\odot}-75\,\text{km}$
and $r_{2}=R_{\odot}+200\,\text{km}$. We find a reasonably good match,
verifying the validity of the reciprocity relation.

\section{Three-dimensional kernel\label{sec:3D-kernel}}

A change in cross-covariance of the radial components of displacement
may be expressed as 
\begin{align}
\delta C_{rr}\left(\mathbf{x}_{1},\mathbf{x}_{2};\omega\right) & =\omega^{2}P\left(\omega\right)\int d\Omega_{\text{src}}\,\left[\delta G_{rr}^{*}\left(\mathbf{x}_{1},\mathbf{x}_{\text{src}};\omega\right)G_{rr}\left(\mathbf{x}_{2},\mathbf{x}_{\text{src}};\omega\right)+G_{rr}^{*}\left(\mathbf{x}_{1},\mathbf{x}_{\text{src}};\omega\right)\delta G_{rr}\left(\mathbf{x}_{2},\mathbf{x}_{\text{src}};\omega\right)\right],\label{eq:dCrr_defn}
\end{align}
where the radial component of the Green function is obtained from
Equation \eqref{eq:divG_HansenVSH} to be
\[
G_{rr}\left(\mathbf{x}_{i},\mathbf{x}_{\text{src}};\omega\right)=\sum_{j}\frac{\left(2j+1\right)}{4\pi}G_{rrj\omega}\left(r_{i},r_{\text{src}}\right)P_{j}\left(\hat{n}_{i}\cdot\hat{n}_{\text{src}}\right),
\]
and the change in the radial component due to a local sound-speed
inhomogeneity $\delta c\left(\mathbf{x}\right)$ may be represented
as 
\[
\delta G_{rr}\left(\mathbf{x}_{\text{obs}},\mathbf{x}_{\text{src}},\omega\right)=-2\int d\mathbf{x}\,\rho c\grad\cdot\mathbf{G}_{r}\left(\mathbf{x},\mathbf{x}_{\text{obs}},\omega\right)\grad\cdot\mathbf{G}_{r}\left(\mathbf{x},\mathbf{x}_{\text{src}},\omega\right)\delta c\left(\mathbf{x}\right).
\]
The radial component of the divergence may be calculated to be 
\[
\grad\cdot\mathbf{G}_{r}\left(\mathbf{x},\mathbf{x}_{i};\omega\right)=\sum_{j}\frac{\left(2j+1\right)}{4\pi}\left[\grad\cdot\mathbf{G}\right]_{rj\omega}\left(r,r_{i}\right)P_{j}\left(\hat{n}\cdot\hat{n}_{i}\right),
\]
where the radial profile $\left[\grad\cdot\mathbf{G}\right]_{rj\omega}\left(r,r_{i}\right)$
is given by the $\beta=0$ term in Equation \eqref{eq:divG_radial}.
The kernel is defined through the equation 
\begin{equation}
d\tau\left(\mathbf{x}_{1},\mathbf{x}_{2}\right)=\int d\mathbf{x}\,K\left(\mathbf{x},\mathbf{x}_{1},\mathbf{x}_{2}\right)\delta c\left(\mathbf{x}\right).\label{eq:kernel_3D_definition}
\end{equation}
Substituting Equation \eqref{eq:dCrr_defn} into Equation \eqref{eq:dtau_dC}
and recasting it into the form of Equation \eqref{eq:kernel_3D_definition},
we obtain 
\begin{align*}
K\left(\mathbf{x},\mathbf{x}_{1},\mathbf{x}_{2}\right) & =\int_{0}^{\infty}\frac{d\omega}{2\pi}\,\omega^{2}P\left(\omega\right)\sum_{j_{1}j_{2}}\frac{\left(2j_{1}+1\right)}{4\pi}\frac{\left(2j_{2}+1\right)}{4\pi}P_{j_{1}}\left(\hat{n}\cdot\hat{n}_{1}\right)P_{j_{2}}\left(\hat{n}\cdot\hat{n}_{2}\right)\times\\
 & 2\Re\left[h^{*}\left(\mathbf{x}_{1},\mathbf{x}_{2},\omega\right)T_{0,j_{1}j_{2}\omega}^{0}\left(r,r_{1},r_{2},r_{\text{src}}\right)\right].
\end{align*}
We simplify the angular function
\[
A=\frac{\left(2j_{1}+1\right)}{4\pi}\frac{\left(2j_{2}+1\right)}{4\pi}P_{j_{1}}\left(\hat{n}\cdot\hat{n}_{1}\right)P_{j_{2}}\left(\hat{n}\cdot\hat{n}_{2}\right)
\]
by using rewriting the expression in terms of spherical harmonics.
We use the expansion of Legendre polynomials in terms of spherical
harmonics:
\[
\frac{\left(2j+1\right)}{4\pi}P_{j}\left(\hat{n}\cdot\hat{n}_{i}\right)=\sum_{m=-j}^{j}Y_{jm}\left(\hat{n}\right)Y_{jm}^{*}\left(\hat{n}_{i}\right)
\]
to obtain the angular term to be 
\[
A=\sum_{m_{1}=-j_{1}}^{j_{1}}\sum_{m_{2}=-j_{2}}^{j_{2}}Y_{j_{1}m_{1}}\left(\hat{n}\right)Y_{j_{2}m_{2}}\left(\hat{n}\right)Y_{j_{1}m_{1}}^{*}\left(\hat{n}_{1}\right)Y_{j_{2}m_{2}}^{*}\left(\hat{n}_{2}\right).
\]
We use the angular momentum coupling relation 
\[
Y_{j_{1}m_{1}}\left(\hat{n}\right)Y_{j_{2}m_{2}}\left(\hat{n}\right)=\sum_{jm}\sqrt{\frac{\left(2j_{1}+1\right)\left(2j_{2}+1\right)}{4\pi\left(2j+1\right)}}C_{j_{1}0j_{2}0}^{j0}C_{j_{1}m_{1}j_{2}m_{2}}^{jm}Y_{jm}\left(\hat{n}\right)
\]
\citep[see][]{1988qtam.book.....V} to obtain 
\[
A=\sum_{jm}\sqrt{\frac{\left(2j_{1}+1\right)\left(2j_{2}+1\right)}{4\pi\left(2j+1\right)}}C_{j_{1}0j_{2}0}^{j0}\sgshj{j_{1}j_{2}*}\left(\hat{n}_{1},\hat{n}_{2}\right)Y_{jm}\left(\hat{n}\right).
\]
We identify the constant pre-factor to be $N_{j_{1}j_{2}j}$ from
Equation \eqref{eq:Nj1j2l}.

\bibliographystyle{aasjournal}
\bibliography{references}

\end{document}